\documentclass[twocolumn,prb,aps,amsmath,amssymb,superscriptaddress,showpacs]{revtex4}
\usepackage{graphicx}
\usepackage{bm}
\usepackage[usenames,dvipsnames]{color}
\usepackage{autobreak}

\renewcommand\Re{\operatorname{Re}}

\newcommand{\bra}{\langle}
\newcommand{\ket}{\rangle}

\newcommand\End{\mathrm{End}}
\newcommand\id{\mathrm{Id}}
\newcommand\Tr{\mathrm{Tr}}
\newcommand{\tr}{\mathrm{Tr}}
\newcommand{\trans}{\mathrm{T}}

\newcommand\calA{\mathcal{A}}
\newcommand\calB{\mathcal{B}}

\newcommand\calD{\mathcal{D}}

\newcommand\calJ{\mathcal{J}}
\newcommand\calK{\mathcal{K}}
\newcommand\calL{\mathcal{L}}
\newcommand\calM{\mathcal{M}}
\newcommand\calO{\mathcal{O}}
\newcommand\calS{\mathcal{S}}
\newcommand\calT{\mathcal{T}}
\newcommand\calY{\mathcal{Y}}
\newcommand\bbC{\mathbb{C}}
\newcommand\bbH{\mathbb{H}}

\newcommand{\barL}{{\overline{L}}}
\newcommand{\barR}{{\overline{R}}}

\newcommand{\ZZ}{{\mathbb Z}}
\begin{document}
\title{Noncommutative geometry,  the Lorentzian Standard Model and its B-L extension}

\author{Fabien Besnard}
 \affiliation{P{\^o}le de recherche M.L. Paris,
 EPF, 3~bis rue Lakanal,
 F-92330 Sceaux, France}

\author{Christian Brouder}
\affiliation{Sorbonne Universit\'e, 
CNRS UMR 7590,
Mus\'eum National d'Histoire
Naturelle,\\
Institut de Min\'eralogie de Physique des Mat\'eriaux et de
Cosmochimie, 
IMPMC, F-75005 Paris, France.
}

\date{\today}
\begin{abstract}
We explore the 1-loop renormalization group flow of two models coming from a generalization of the Connes-Lott version of Noncommutative Geometry in Lorentzian signature: the Noncommutative Standard Model and its B-L extension.  Both make predictions on coupling constants at high energy, but only the latter is found to be compatible with the top quark and Higgs boson masses at the electroweak scale. We took into account corrections introduced by threshold effects and the relative positions of the Dirac and Majorana neutrino mass matrices and found them to be important. Some effects of 2-loop corrections are briefly discussed. The model is consistent with experiments only for a very small part of its parameter space and is thus predictive. The masses of the $Z'$ and B-L breaking scalar are found to be of the order $10^{14}$ GeV. 
\end{abstract}
\maketitle
 

\section{Introduction}
Noncommutative geometry (NCG) is a remarkably elegant mathematical
framework {which allows} to derive the {field content} and Lagrangian of the Standard Model
of particle physics~\cite{Connes-Marcolli,Suijlekom}.
The history of the NCG approach to the Standard Model (SM) is 
described in a recent paper~\cite{Chamseddine-19}. A first landmark is Ref.~\onlinecite{Connes-90} where Connes and Lott obtained  
the SM  {bosonic} Lagrangian {thanks to a universal formula of Yang-Mills type, i.e. the squared length of the curvature of 
a single non-commutative one-form which encapsulates simultaneously the gauge and Higgs fields}. This construction
automatically generates the {quartic potential for the} Higgs field. 
In 1995, Connes\cite{Connes-95-reality} added a key element to his construction, namely a charge reversal
operator (i.e. a real structure). In 1996, Chamseddine and Connes\cite{Connes-96}
observed that the SM {bosonic} Lagrangian can be obtained directly
by using the ``spectral action principle'': the physical action depends
only on the spectrum of the Dirac operator.  {This is a major breakthrough since the Einstein-Hilbert action evaluated on the manifold metric turns out to be a component of the spectral action. This raises the hope of understanding all the known fundamental forces as the different facets of a unique gravitational field defined on a generalized manifold, fulfulling the dream of Einstein, Kaluza and Klein.}  Right-handed neutrinos 
were added in 2006\cite{Connes-Chamseddine-06}: the type I see-saw mechanism is then automatically triggered by the general principles of NCG. When recovering the SM bosonic Lagrangian through the spectral action, relations between couplings are obtained. In particular the gauge couplings are unified, which means that these relations hold only at high energy. When running  the renormalization group equations down to the electroweak scale, one obtains a prediction for the top quark and Higgs boson masses.  

The spectral action has a physical drawback: to date, it has only been possible to define it for Euclidean spacetimes. After this action is evaluated, it is then necessary to perform a Wick rotation.  {On the other hand, the Connes-Lott action can be defined on Lorentzian spacetimes, as shown by Karen Elsner~\cite{Elsner-99}.} In this paper, we insist on using the physically correct signature right from the start, and this is why we will use the second approach. It is admitedly less ambitious: by using a noncommutative 1-form as the bosonic variable, the Connes-Lott approach is a noncommutative version of gauge theory, which forgets about gravity, wheras the Connes-Chamseddine spectral action uses directly the metric, in close analogy with Kaluza-Klein theory. However, it must be said that in order to promote Connes-Chamseddine theory to a full-fledged noncommutative Kaluza-Klein theory, one has to define a structure in which the Dirac operator may vary, and would be to spectral triples what bare differentiable manifolds are to Riemannian manifolds. Such a structure has   been recently proposed\cite{Besnard-19-2} in the form of \emph{algebraic backgrounds}.

In a previous article~\cite{Bizi-18}, we put forward a  {working definition of }\emph{indefinite spectral triples}, applicable to 
pseudo-Riemannian manifolds and taking charge conjugation
and parity into account.
We showed that, in this approach, the Lagrangian of 
quantum electrodynamics in Lorentzian spacetime was recovered. In the present work we build an indefinite spectral triple
to recover the Lagrangian of the Standard Model (including
right-handed neutrinos) on a smooth globally hyperbolic
(Lorentzian) four-dimensional manifold $M$,
which we assume time- and space-orientable.   In order to do this, we will introduce in section \ref{SRNC} the strictly necessary material on indefinite spectral triples and algebraic backgrounds, without requiring any previous knowledge of the subject. We will then be able to define these particular structures in the case of the SM, in section \ref{SMback}. The next step will be a necessary updating of   NC gauge theory to take into account the real structure and semi-Riemannian signature (section \ref{NCGT}). When this is done, we will apply these tools to the indefinite spectral triple of the SM defined earlier and obtain in section \ref{NCSM} a particle model which has exactly the same field content and Lagrangian as the SM extended by right-neutrinos and type I see-saw, \emph{except} 1) it has an additional abelian gauge field $X$, and 2) there are constraints on the parameters of the Lagrangian, such as the unification of the gauge couplings. Once the $X$-field is removed by the unimodularity condition (as is usual in NCG), we   obtain the SM with the correct physical signature entirely within the framework of NCG\footnote{This  result, up to a few details, has been first obtained by Nadir Bizi in Ref.~\onlinecite{Bizi-PhD}.}. Running down the renormalization group equations (RGE) from some unification energy scale $\mu_{\rm unif}$ (which is a free parameter), we confirm in this new context the result already obtained with the spectral action\cite{Chamseddine-06}: the predicted Higgs mass is at least $30\%$ too large. It was observed in Ref.~\onlinecite{Chamseddine-12} that one can remedy this situation by adding a real scalar field to the model. We will briefly review the results of renormalization group analysis of this scalar extended model in section \ref{CCmodel}, also including threshold corrections which were not taken into account in Ref.~ \onlinecite{Chamseddine-12}. It has been an important trend in later years to find theoretical motivations for this new scalar\cite{Boyle-14,Boyle-16,Devastato-14,Devastato-15}. The most direct explanation from a theoretical point of view is probably the embedding of the model into a noncommutative version of Pati-Salam theory\cite{Chamseddine-13-PS}, but this requires important modifications to the usual formalism of NCG\cite{Chamseddine-13}. Moreover the RGE of the full Pati-Salam model are quite involved and their analysis relies on many assumptions\cite{Aydemir-16,Chamseddine-16} . We will propose instead in section \ref{BLextSM} a simpler extension of the noncommutative SM. This follows from the observation that in the    framework of algebraic backgrounds, the configuration space of NC gauge theory is a subspace of the configuration space of NC Kaluza-Klein theory which is stable under certain symmetries. In the case of the SM background, these symmetries include the B-L gauge symmetries, as shown in Ref.~\onlinecite{Besnard-19-3}. Hence, from the algebraic background point of view, the noncommutative SM is not a consistent theory: only its B-L extension is. Finally we perform the RG analysis of the B-L extended SM with the initial conditions yielded by the Connes-Lott-Elsner action. We show that there are only 3 relevant parameters $n$, $\rho$ and $\epsilon$, which are in order a normalization constant, the quotient of two Yukawa couplings, and the angle between the Dirac and Majorana mass matrices for neutrinos. We find that there exist a region of the parameter space which gives   good fit  to the experimental values. Using bounds on light neutrino masses, we observe that the model predicts very high ($\ge 10^{14}$ GeV) masses for the $Z'$ bosons and the B-L-breaking Higgs.

If one is ready to take the mathematical motivations for granted, the reading can start directly at section \ref{NCSM}.

\section{The semi-Riemannian NCG framework} \label{SRNC}

The main difference between our indefinite spectral triple
(see next section for a precise definition) and the spectral
triples of Euclidean NCG is the
   replacement of the Hilbert space by a pre-Krein space, i.e. a complex vector space equipped with a Hermitian form $(.,.)$ such that $(\psi,\psi)$ can be of arbitrary sign depending on the vector $\psi$.  Indeed, working with a non-compact Lorentzian manifold instead of a compact Riemannian one triggers a cascade of effects that need to be taken into account. First, the metric on spinor fields becomes indefinite (through the insertion of a $\gamma^0$ matrix). This means that the smooth part of the almost-commutative spectral triple under construction will cease to be Euclidean. Less obviously, this feature will be transmitted to the  finite-dimensional part through the general rules for forming tensor products of indefinite spectral triples. Even less obviously perhaps, the completion of the space of spinor fields ceases to be unique because of the non-compactness of the base manifold \cite{Besnard-18-3}, forcing us to use the   pre-Krein space of  compactly supported fields instead of an arbitrary $L^2$-completion.

\subsection{Indefinite spectral triples}
For the convenience of the reader, we repeat the definition
of an indefinite spectral triple that we recently
put forward~\cite{Bizi-18}.  The only difference is that we put the Krein product to the forefront instead of the fundamental symmetry. We explain this choice below. We recall that a Krein product is a  non-degenerate Hermitian form. The adjoint of an operator $T$ with respect to a Krein product will be denoted by $T^\times$, except when the Krein product is positive-definite, in which case we revert to the more traditional $T^\dagger$. 

An indefinite real even spectral triple (IST) is a tuple $(\calA,\calK, \pi,\chi,J,D)$
where
\begin{enumerate}
\item $\calA$ is a real or complex *-algebra,
\item $\calK$ is a complex {pre}-Krein space,
\item $\pi$ is  $*$-representation of $\calA$ on $\calK$,
\item   $\chi$ (chirality) is a linear operator on $\calK$ and $J$ (real structure or charge conjugation) is an antilinear operator. It is required that  $[\chi,\pi(a)]=0$ for all $a\in\calA$,   $\chi^2=1$ and
     \begin{align}
     J^2=\epsilon,\ J\chi=\epsilon''\chi J\cr
	J^\times=\kappa J,\ \chi^\times=\epsilon''\kappa''\chi,\label{kosigns}
	\end{align}      
     where $\epsilon,\epsilon'',\kappa,\kappa''$ are signs, 
 
\item  $D$ (the Dirac) is an operator  on $\calK$ which commutes with $J$,
  anticommutes with $\chi$ and satisfies
  $(D\psi,\psi') =(\psi,D\psi')$ for all $\psi,\psi'$ in its domain.
\end{enumerate}

Note that the above definition is only the  core of a more complete set of axioms which is still under construction. In particular we refer the reader to
Refs.~\onlinecite{Strohmaier-06,Dungen-13,Franco-14-2,Dungen-16} for the
functional analytic conditions satisfied by $\pi$ and $D$  (in the context of a fixed fundamental symmetry and Hilbert completion).

It is common to add some other properties:
i) the condition of order zero (i.e. for any
$a$ and $b$ in $\calA$, $\pi(a)$ commutes with
$\pi(b)^\circ:=J \pi(b)^\times J^{-1}$);
ii) the condition of order one (i.e. for any
$a$ and $b$ in $\calA$, $[D,\pi(a)]$ commutes with
$\pi(b)^\circ$).  In this paper we will assume the order-zero condition  only ($C_0$).

A \emph{fundamental symmetry} $\eta$ on an IST is an operator  which either commutes or anticommutes with $\chi$ and $J$, squares to $1$, and  combine with the Krein product $(.,.)$  to form  a scalar product $\bra.,.\ket_\eta:=( .,\eta.)$. Such a fundamental symmetry always exists \cite{Besnard-19-2} (but is far from unique) and given one the second line of \eqref{kosigns}  is equivalent to
\begin{equation}
J\eta=\epsilon\kappa \eta J,\ \chi\eta=\epsilon''\kappa'' \eta\chi
\end{equation}

The four signs $\epsilon,\epsilon'',\kappa,\kappa''$ depend on two even integers $m$ and $n$, unique
modulo 8, such that 
$\epsilon=(-1)^{n(n+2)/8}$,
$\epsilon''=(-1)^{n/2}$,
$\kappa=(-1)^{m(m+2)/8}$ and
$\kappa''=(-1)^{m/2}$. Integer $n$ is the usual 
$KO$ dimension of NCG, integer $m$, called
the \emph{metric dimension}, is an additional integer
required to classify indefinite spectral triples\cite{Bizi-18}. Note that when $\epsilon''\kappa''=1$ (resp. $-1$), the two eigenspaces of $\chi$ are orthogonal with one another (resp. self-orthogonal) and  fundamental symmetries must commute (resp. anticommute) with $\chi$. In that case we say that the Krein product is even (resp. odd). This will play a role below in the definition of tensor products.

{\small {\bf Remark 1:} We need to make a comment on the adjoint of an anti-linear operator. It is defined by 
\begin{eqnarray}
(\phi,A\psi) &=&s (\psi,A^\times\phi).
\label{standard_dual}
\end{eqnarray}
where $s=1$ if $\phi$ and $\psi$ are treated as a commuting variables and $s=-1$ if they are treated as anti-commuting ones. Since NCG is a classical theory, both cases can be found in the literature. However the choice $s=1$ would make things go astray at several places (in the definition of the Lagrangian, for the see-saw mechanism and for solving the fermion doubling). We will thus consider only the $s=-1$ case in this paper.}

{\small {\bf Remark 2:} It would be tempting to consider the Hilbert completion of $\calK$ with respect to $\eta$ and formulate the theory in terms of a Hilbert space and a fundamental symmetry instead of the less familiar (pre-)Krein space. However it would be a bad idea for two reasons. The first is that this completion is generally not unique \cite{Besnard-18-3}, but even when it is, for instance in finite-dimension, we would then put two objects (the scalar product and the fundamental symmetry)  instead of one (the Krein product) in the background, which would pose a conceptual problem for the definition of symmetries (the example of Minkowski space where choosing $\eta$ is equivalent to choosing a time coordinate is good to keep in mind).}

\subsection{Noncommutative $1$-forms}
The theory of noncommutative 1-forms exposed in Ref.~\onlinecite{Connes94} (to which we refer for more details) can be extended without change\footnote{The only issue that can arise is the existence of the projection operator on junk forms represented on the pre-Krein space. We will see in section \ref{NCGT} that this problem does not show up in the models we consider in this paper.} to the indefinite setting. We recall here the main concepts. Let $\calS=(\calA,\ldots,D)$ be an IST.  An element $\omega\in {\rm End}(\calK)$ of the form 
\begin{equation}
\omega=\sum_i \pi(a_i)[D,\pi(b_i)], a_i,b_i\in\calA\label{def1forms}
\end{equation}
is called a \emph{noncommutative $1$-form} of $\calS$. The space of such forms is written $\Omega^1_D$. It is a bimodule over $\calA$, and the map $d_D: a\mapsto [D,\pi(a)]$ is a derivation of $\calA$ into $\Omega^1_D$  which is a first-order differential calculus in the sense of Ref.~\onlinecite{Woronowicz-89}. One  extends $d_D$ to $\Omega^1_D$ by
\begin{equation}
d_D\omega=\sum_i[D,\pi(a_i)][D,\pi(b_i)].\label{defder}
\end{equation}
However, $\omega$ can be decomposed as in \eqref{def1forms} in several ways, hence \eqref{defder} makes sense modulo a certain ideal  $\calJ^1_D$ of so-called\cite{Iochum-95}   ``junk $2$-forms''. The curvature of a $1$-form  is defined modulo junk by 
\begin{equation}
\rho_D(\omega)=d_D\omega+\omega^2.\label{defcurv}
\end{equation}
Let $u$ be an invertible element of $\calA$. It defines a gauge transformation on $1$-forms by the formula
\begin{equation}
\omega\mapsto \omega^u:=\pi(u)\omega \pi(u)^{-1}+\pi(u)[D,\pi(u)^{-1}].\label{gaugetransf}
\end{equation}
Then the curvature is gauge-covariant:
\begin{equation}
\rho_D(\omega^u)=u\rho_D(\omega)u^{-1}.\label{gaugecov}
\end{equation}
\subsection{Algebraic backgrounds}
For applications to physics it is important to define a background structure in which the Dirac operator can vary. For several reasons\cite{Besnard-19-2,Besnard-19-3,Besnard-20-1} one cannot just remove the Dirac operator from a spectral triple. Instead we define an \emph{algebraic background} to be a tuple $\calB=(\calA,\ldots,J,\Omega^1)$, where the objects $\calA,\ldots,J$ are exactly the same as in the definition of an IST, and $\Omega^1$ is an odd $\calA$-bimodule  (its elements anticommute with $\chi$).

The symmetries of a background $\calB$ are naturally defined to be the Krein unitary operators $U$ which commute with $\chi$ and $J$, and stabilize $\pi(\calA)$ and $\Omega^1$. A particularly important case is the following. Let $u$ be a unitary element of $\calA$ and define the \emph{gauge transformations} $\Upsilon(u):=\pi(u)\pi(u^{-1})^o$. These transformations will be symmetries of $\calB$ under the condition
\begin{equation}
\pi(u^{-1})^o\Omega^1\pi(u)^o=\Omega^1,\label{weakC1}
\end{equation}
for all unitary $u$. We call \eqref{weakC1} the \emph{weak order one condition} (weak $C_1$).  Clearly the usual order one condition implies the weak one.

An operator $D$ is called a \emph{compatible Dirac operator} for $\calB$ if it has all the properties of a Dirac operator listed in the definition of an IST and satisfies $\Omega^1_D\subset \Omega^1$. It is moreover called \emph{regular} if $\Omega^1_D= \Omega^1$. We will always suppose that at least one regular Dirac exists. The vector space of all compatible Dirac operators for $\calB$ is called the \emph{configuration space}. It is stable by the symmetries of $\calB$.

Given a compatible Dirac $D$ and a selfadjoint $1$-form $\omega$, one defines the \emph{fluctuated Dirac operator}
\begin{equation}
D_\omega=D+\omega+\omega^o,\label{fluct}
\end{equation}
where $\omega^o=J\omega^\times J^{-1}$. Let us suppose that for all $a\in\calA$, one has
\begin{equation}
[\omega^o,\pi(a)]\in\Omega^1.\label{weakC1prim}
\end{equation}
We call \eqref{weakC1prim} the \emph{weak $C_1'$ condition}, since it is analogous to weak $C_1$. If this condition holds, it is immediate that $D_\omega$ is also a compatible Dirac.

If weak $C_1$ holds then for every compatible Dirac $D$, $\Upsilon(u)D\Upsilon(u)^{-1}$ is a compatible Dirac, and if in addition $C_1$ holds we have the formula
\begin{eqnarray}
\Upsilon(u)D_\omega\Upsilon(u)^{-1}&=&D_{\omega^u}\label{gt}
\end{eqnarray}
which justifies the name ``gauge transformation'' for \eqref{gaugetransf}.

\subsection{Tensor products}
The general rules for the tensor product of two IST are the following ones \cite{Bizi-PhD,Bizi-18}. 

Let  $\calB_1=(\calA_1,\calK_1,\ldots,J_1,\Omega^1_1)$ and $\calB_2=(\calA_2,\calK_2,\ldots,J_2,\Omega^2_2)$ be two backgrounds. It will be sufficient to consider the case where $\calA_2$ and $\calK_2$ are finite-dimensional. The (graded) tensor product $\calB=\calB_1\hat\otimes \calB_2:=(\calA,\calK,\ldots,J,\Omega^1)$ is defined in the following way. First we set $\calA=\calA_1\otimes\calA_2$, $\calK=\calK_1\otimes \calK_2$, $\pi=\pi_1\otimes \pi_2$, $\chi=\chi_1\otimes \chi_2$. In order to define the rest of the structure, let us define some notation. If $\psi$ is in one of the eigenspaces of $\chi$ we say that it is homogeneous, and we define its grading $|\psi|\in\ZZ_2$ to be equal to $0$ if $\chi\psi=\psi$ and $1$ if $\chi\psi=-\psi$. Similarly,  operators commuting  with the chirality are said to be \emph{even} and given the grading $0$, while operators anti-commuting with it are said to be \emph{odd} and given the grading $1$.  For homogeneous operators $T_i\in \End(\calK_i)$, $i=1,2$, we can define the   graded tensor product $T_1\hat\otimes T_2$ by
\begin{equation}
(T_1\hat\otimes T_2)(\psi_1\otimes \psi_2):=(-1)^{|\psi_1||T_2|}T_1\psi_1\otimes T_2\psi_2.
\end{equation}
The graded tensor product of homogeneous operators is related to the usual tensor product by the formula $T_1\hat\otimes T_2=T_1\chi_1^{|T_2|}\otimes T_2$. With these notations in hand we define the real structure $J$ to be
\begin{equation}
J_1\chi_1^{|J_2|}\hat\otimes J_2\chi_2^{|J_2|}.
\end{equation}
The bimodule $\Omega^1$ will be generated by the $1$-forms
\begin{equation}
\omega=\omega_1\hat\otimes 1+1\hat\otimes \omega_2,\ \omega_1\in \Omega^1_1,\omega_2\in\Omega^1_2
\end{equation}
The Krein product on $\calK$  is defined by
\begin{equation}
(\phi_1\hat\otimes \phi_2,\psi_1\hat\otimes \psi_2)=(\phi_1,\psi_1)_1(\phi_2,\beta \psi_2)_2,\label{tenskprod}
\end{equation}
where $\beta=1$ if $(.,.)_1$ is even, $\beta=\chi_2$ if $(.,.)_1$ is odd and $(.,.)_2$ is even, and $\beta=i\chi_2$ if $(.,.)_{1,2}$ are both odd. Note that the KO and metric dimensions are additive with respect to tensor products. Finally we observe that if $D_1,D_2$ are compatible (resp. regular) Dirac operators for $\calB_1,\calB_2$ respectively, then
\begin{equation}
D=D_1\hat\otimes 1+1\hat\otimes D_2,\label{tensD}
\end{equation}
is a compatible (resp. regular) Dirac operator for $\calB$. Consequently the tensor product of two IST $\calS_1=(\calA_1,\ldots,J_1,D_1)$ and $\calS_2=(\calA_2,\ldots,J_2,D_2)$ is defined by $\calS=(\calA,\ldots,J,D)$, with $\calA,\ldots,J$  as above and $D$ given by \eqref{tensD}.

\section{SM algebraic background and IST}\label{SMback}
The IST adapted to the Standard Model
is very close to the spectral triple defined by Connes and coll.~\cite{Connes-Marcolli},
except for the fact that we work in a Lorentzian four-dimensional spacetime
$M$ (signature $(1,3)$). First, out of $M$ we build the background $\calB_M=(\calA_M,\ldots,\Omega^1_M)$   where $\calA_M=\tilde C^\infty(M)_c$ is the algebra of 
real-valued smooth functions over $M$ which are constant outside a compact (this is the  unitization of the algebra of compactly supported functions), $\calK_M$ is the  space of compactly supported  
spinor fields, $\pi_M$ is the representation of functions by multiplication on spinors, $\chi_M$ is the multiplication by $\gamma_5$, the Krein product is\footnote{The formula is given at a point. For a spinor field, integrate over the manifold.} $(\psi,\psi')=\psi^\dagger\gamma_0\psi'$, $J_M \psi=\gamma_2\psi^*$. Finally the $1$-forms are just the usual $1$-forms represented on $\calK_M$ by Clifford multiplication. A regular Dirac operator for $M$ is the canonical Dirac operator\footnote{We consider the flat case here. For the general case, replace partial derivative with the canonical spin connection.} $D_M=i\gamma^\mu\partial_\mu$. Note that the KO-metric pair is $(6,4)$ so that $\epsilon=1,\epsilon''=-1,\kappa=-1,\kappa''=1$. The   background $\calB_M$ and the IST $\calS_M=(\calA_M,\ldots,D_M)$ are respectively called the canonical background and IST of $M$, respectively.

The IST of the Standard Model is $\calS_{SM}:=\calS_M\hat\otimes \calS_F$ where $\calS_F$ is a finite IST that we now need to describe. The algebra is $\calA_F=\bbC\oplus \bbH \oplus M_3(\bbC)$, where
$\bbH$ is the algebra of quaternions. The   Krein space is 
\begin{eqnarray}
\calK_F&=&\calK_R\oplus \calK_L \oplus \calK_\barR\oplus\calK_\barL\label{KF},
\end{eqnarray}
where these four  spaces represent the right particles, left particles,
anti-right-particles and anti-left-particles. 
Each $\calK_i$ is 24-dimensional and isomorphic to
\begin{eqnarray}
\calK_0 &=& (\bbC_\ell^2\,\oplus\, \bbC_q^2\otimes\bbC_c^3)\otimes 
\bbC_g^N.\label{K0}
\end{eqnarray}
The relation with the physical particles is the following
\begin{itemize}
\item $\bbC_\ell^2$ is a lepton doublet of  {canonical} basis $(\nu,e)$
\item $\bbC_q^2$ is a quark doublet of  {canonical} basis $(u,d)$
\item $\bbC_c^3$ is the space of colours $(r,g,b)$ or $(1,2,3)$
\item $\bbC_g^N$ is the space of generations (usually $N=3$)
\end{itemize}
For example, a basis of the space $\calK_R$ of right particles is made of
$(\nu_R,e_R,u^r_R,u^g_R,u^b_R,d^r_R,d^g_R,d^b_R)$ for each generation,
a basis of the space $\calL_{\bar L}$ of anti-left-particles (which are right-handed antiparticles) is made of
$(\nu_L^c,e_L^c,u^{rc}_L,u^{gc}_L,u^{bc}_L,
d^{rc}_L,d^{gc}_L,d^{bc}_L)$ for each generation. Another way to look at \eqref{K0} is to see $\ell$ as the fourth colour as in Pati-Salam theory, and write 
\begin{eqnarray}
\calK_0 &=& \bbC^2_I\otimes \bbC^4_c\otimes 
\bbC_g^N. 
\end{eqnarray}
With this decomposition we can introduce the useful notation $\tilde a:=a\otimes 1_4\otimes 1_N$. Using this notation, the representation $\pi_F$ is defined as follows:  for an element $(\lambda,q,m)\in \bbC\oplus\bbH\oplus M_3(\bbC)$, one defines
\begin{equation}
\pi_F(\lambda,q,a)={\rm diag}(\tilde q_\lambda,\tilde q,1_2\otimes(\lambda\oplus a)\otimes 1_N, 1_2\otimes(\lambda\oplus a)\otimes 1_N)\label{SMrep}
\end{equation} 
where $q_\lambda=\begin{pmatrix}\lambda&0\cr 0&\lambda^*\end{pmatrix}$ is the embedding of $\bbC$ into $\bbH$ seen as the algebra of matrices of the form $\begin{pmatrix}\alpha&\beta\cr -\beta^*& \alpha^*\end{pmatrix}$ and $\lambda\oplus a$ is the block diagonal matrix $\begin{pmatrix}\lambda&0\cr 0&a\end{pmatrix}$ acting on the colour $\bbC^4$. Moreover \eqref{SMrep} is seen as  a diagonal matrix in the decomposition \eqref{KF}. Using the same decomposition (which we will use hereafter without further notice), the chirality operator is
\begin{equation}
\chi_F={\rm diag}(1,-1,-1,1)\label{SMchirality}
\end{equation}
where $1$ is the identity operator on $\calK_0$. The Krein product on $\calK_F$ is $(.,.)_F=\bra.,\eta_F.\ket$, with fundamentaly symmetry
\begin{equation}
\eta_F={\rm diag}(1,-1,-1,1)=\chi_F.\label{etaF}
\end{equation}

The real structure is
\begin{equation}
J_F=\begin{pmatrix}0&0&-1&0\cr 0&0&0&-1\cr 1&0&0&0\cr 0&1&0&0 \end{pmatrix} \circ c.c.\label{JF}
\end{equation}
with the same notation and c.c. means complex conjugation. The finite Dirac is
\begin{equation}
D_F=\begin{pmatrix}
0&-\Upsilon^\dagger&-M^\dagger&0\cr \Upsilon&0&0&0\cr M&0&0&-\Upsilon^T\cr 0&0&\Upsilon^*&0
\end{pmatrix},\label{DF}
\end{equation}
where 
\begin{equation}
\Upsilon=\begin{pmatrix}\Upsilon_\ell&0\cr 0&\Upsilon_q\otimes 1_3\end{pmatrix},\label{Y}
\end{equation}
with $\Upsilon_\ell,\Upsilon_q\in M_2(M_N(\bbC))$ given by
\begin{equation}
\Upsilon_\ell=\begin{pmatrix}\Upsilon_\nu&0\cr 0&\Upsilon_e\end{pmatrix},\ \Upsilon_q=\begin{pmatrix}\Upsilon_u&0\cr 0&\Upsilon_d\end{pmatrix},\label{formeYell}
\end{equation}
where we have decomposed the $\bbC^2$ factor using the $(u,d)$ basis, while
\begin{equation}
M= \begin{pmatrix}m&0\cr 0&0\end{pmatrix}\otimes\begin{pmatrix}1&0&0&0\cr 0&0&0&0\cr 0&0&0&0\cr 0&0&0&0\end{pmatrix}\label{formeM}
\end{equation}
where $m\in   M_N(\bbC)$  is  a symmetric   matrix (responsible for the type I see-saw mechanism). This ends the definition of $\calS_{SM}$.

The SM background $\calB_{SM}$ is the tensor product $\calB_M\hat\otimes\calB_F$, where $\calB_F$ is the finite background constructed out of the same objects as $\calS_F$ except that we replace $D_F$ with $\Omega^1_F:=\Omega^1_{D_F}$ (so that $D_F$ is a regular Dirac by construction).  The bimodule $\Omega^1_F$ contains matrices of the form
\begin{equation}
\omega_F=\begin{pmatrix}
0&\Upsilon^\dagger \tilde q_1&0&0\cr \tilde q_2 \Upsilon&0&0&0\cr 0&0&0&0\cr 0&0&0&0
\end{pmatrix}, q_1,q_2\in \bbH.
\end{equation}
The definitions of the SM triple and background seem extremely contrived, but the beauty of the NCG approach is that there is on the contrary very little freedom in these choices. Clearly definition \eqref{K0} is dictated by the fermionic content of the theory, while the choice of the algebra is motivated by the gauge group. One can see from \eqref{KF} a quadruplication of the fermionic degrees of freedom, since $\calK_M$ already contains four-dimensional Dirac spinor fields. This problem, known for a long time\cite{Lizzi-97}, is solved by defining the physical Krein space by the Majorana-Weyl conditions\cite{Barrett-07}
\begin{eqnarray}
J\Psi&=&\Psi,\cr
\chi\Psi&=&\Psi.\label{MWconditions}
\end{eqnarray}
This solution can be shown to be unique\cite{Besnard-19-4} up to a phase under natural symmetry assumptions, but requires the KO-dimension of the SM background to be $0\ [8]$. Since the KO-dimension of the manifold background is $1-3=6\ [8]$ we obtain that $\calB_F$ has KO-dimension $2\ [8]$. Moreover it can be shown that the fermionic action is non-vanishing only if the metric dimension of $\calB_{SM}$ is $2\ [8]$, which yields a metric dimension of $6\ [8]$ for the finite background. These constraints completely determine \eqref{SMchirality},\eqref{etaF},\eqref{JF} up to a change of basis. The Dirac operator $D_F$ is also strongly constrained by the IST axioms as well as the order $1$ condition. The forms \eqref{DF} and \eqref{Y} are the most general, while there exist other solutions beyond \eqref{formeYell},\eqref{formeM} which are here taken to be the simplest  non-trivial ones. There are some theoretical arguments to reduce the freedom even more \cite{Boyle-14,Brouder-15-NCG-1,Dabrowski-18}. It is important to observe in particular that the axioms satisfied by $D_F$ force $m$ to be symmetric.

\section{Noncommutative gauge theory in the presence of a real structure}\label{NCGT}
Noncommutative gauge theory  has been devised by the Connes and Lott at a time when the role of the real structure had not yet come to the forefront. It was also formulated in the Euclidean context. The extension to almost-commutative triples with a manifold part of general signature poses no problem and has already been performed\cite{Elsner-99}. We quickly present here a new version compatible with the presence of $J$ and general signature on the finite part.

Consider a background $\calB=(\calA,\ldots,\Omega^1)$ satisfying the order $0$ condition. Since the fluctuated Dirac in Eq.~\eqref{fluct} contains contributions $\omega$ from $\Omega^1_D$ and
$\omega^\circ$ from $(\Omega^1_D)^\circ$, we use the   \emph{$J$-symmetrized background} $\hat\calB$  obtained by replacing:
\begin{itemize}
\item $\calA$ with the algebra $\hat \calA$ generated by $\pi(\calA)$ and $\pi(\calA)^o$,
\item $\pi$ with $\hat\pi=\id$,
\item $\Omega^1$ with $\hat \Omega^1$, which the $\hat \calA$-bimodule generated by $\Omega^1$ and $(\Omega^1)^o$,
\end{itemize} 
all the other pieces of data remaining unchanged. Note that, using $C_0$, $\hat \calA$ is the image of the  envelopping algebra $\calA\otimes \calA^o$ under $a\otimes b^o\mapsto\pi(a)\pi(b)^o$,where $\calA^\circ$ is the opposite algebra of $\calA$, characterized by 
$a^\circ b^\circ=(ba)^\circ$.  Let $D$ be a regular Dirac for $\calB$. It is then automatically a regular operator for $\hat\calB$. Let  $\calD_{D}$ be the space of fluctuations \eqref{fluct} of $D$. It is the configuration space of NC gauge theory, and contains all the gauge and Higgs degrees of freedom, while the full configuration space also contains the gravitational degrees of freedom\cite{Besnard-19-2}. We would like to define a gauge-invariant action functional on $\calD_{D}$. This is meaningful if:
\begin{enumerate}
\item\label{a1} gauge transformations are symmetries of $\calB$,
\item\label{a2} $\calD_{D}$ is a subspace of the configuration space of $\calB$,
\item\label{a3} $\calD_{D}$ is gauge-invariant.
\end{enumerate}
All 3 requirements are implied by the order $1$ condition which holds for the SM.  In the B-L-extended SM to be studied below,  weak $C_1$ and $C_1'$ hold, so that requirements  \ref{a1} and \ref{a2} are met. It can be shown\cite{bes-20-b} that \ref{a3}   holds automatically under weak $C_1$. In the B-L case it can also be seen directly or by showing that inner fluctuations  in the sense of Ref.~\onlinecite{Chamseddine-13} are fluctuations in the usual sense \cite{Besnard-19-1}.  For any model satisfying \ref{a1}, \ref{a2}, \ref{a3}, a gauge-invariant action $S(D_\omega)$ can be defined on $\calD_{D}$ by applying any gauge-invariant function to the gauge-covariant curvature $\rho_D(\omega)$ computed in the $J$-symmetrized background. In Connes-Lott theory this function is of Yang-Mills type. In order to be more specific, let us specialize to the case where $\calB=\calB_M\hat\otimes\calB_F$, with $\calB_M$ the canonical background of a manifold and $\calB_F$ a finite-dimensional background. Then we can define the ``Krein-Schmidt product''
\begin{equation}
(A_1,A_2)=\Re\Tr(A_1^\times A_2)
\end{equation}
on operators $A_i$ in ${\rm End}(\bbC^4\otimes \calK_F)$, where $i=1,2$ and $\bbC^4$ is the space of Dirac spinors. Then the \emph{generalized Connes-Lott-Elsner action} is the integral over $M$ of the Lagrangian\footnote{The curvature is evaluated at a point $x\in M$ in this formula.}
\begin{equation}
\calL_b(D_\omega)=-{1\over n}\left(P(\rho_D(\omega)),P(\rho_D(\omega))\right),\label{genCL}
\end{equation}
where $n$ is some constant and $P$ is a projection operator which we now need to describe. We recall that $\rho_D(\omega)$ is only defined modulo the  junk ideal $\hat\calJ^1_D$. The operator $P$ is the projection on the orthogonal of $\hat\calJ^1_D$. Its insertion in \eqref{genCL} makes the formula well-defined. Moreover $P$ has the properties $P=P^\times$, $P(\pi(a)T\pi(b))=\pi(a)P(T)\pi(b)$ from which the reality and gauge-invariance of \eqref{genCL} follow. Note that in a Krein space the orthogonal projection on a subspace $V$ is well defined iff $V\cap V^\perp=\{0\}$, which happens to be the case both for the SM and its B-L-extension. For more details, see Ref.~\onlinecite{Besnard-19-4}.

{\small {\bf Remark:} One can easily prove that 
\begin{equation}
\rho_D(\omega+\omega^o)=\rho_D(\omega)+\rho_D(\omega)^o+\{\omega,\omega^o\},
\end{equation}
and that moreover the term $\{\omega,\omega^o\}$ is in the junk under $C_1$. Using also the property $(A,B^o)=(A^o,B)$ of the Krein-Schmidt product, one can make the dependence of the Connes-Lott-Elsner action on $\omega^o$ disappear entirely. This is the approach followed in Ref.~\onlinecite{Bizi-PhD}. In this case there is no need for $J$-symmetrized background. Since in this paper we will consider the B-L-extension for which $C_1$ does not hold, we must use the most general approach.  Note however that even for the SM there is a subtle difference between the two approaches coming from the fact that the junk ideals $\calJ^1_D$ and $\hat\calJ^1_D$ are not the same.}

\section{The Lagrangian and the RG flow of the NC Standard Model}\label{NCSM}
The Dirac operator around which we fluctuate is $D=D_M\hat\otimes 1+1\hat\otimes D_F$. One can show\cite{Besnard-19-2} that the elements of $\calD_D$ are then of the form
\begin{eqnarray}
D+i\gamma^\mu\hat\otimes (Xt_X+{1\over 2}gB_\mu t_Y+{1\over 2}g_wW^a_\mu t_W^a+{1\over 2}g_sG^a_\mu t_C^a)&&\cr
+1\hat\otimes(\Phi(q-1)+\Phi(q-1)^o)&&\label{Fluctuationdevelopee}
\end{eqnarray} 
where $X,B_\mu,W^a_\mu,G^a_\mu$ are real fields, $q$ is a quaternionic field, $g,g_w,g_s$ are some constants and $t_X,t_Y,t_W^a,t_C^a$ are diagonal matrices of the form ${\rm diag}(\tau_R,\tau_L,\tau_R^*,\tau_L^*)\otimes 1_N$, where in   decomposition \eqref{K0} we have 
\begin{eqnarray*}
\mbox{for }t_X:&&\tau_R=\begin{pmatrix}
0&0\cr 0&-2i \end{pmatrix}\oplus \begin{pmatrix} 0&0\cr 0&-2i\end{pmatrix}\otimes 1_3,\cr
&&\tau_L=-i1_2\oplus -i1_2\otimes 1_3,\cr
\mbox{ for }t_Y:&&\tau_R=\begin{pmatrix}0&0\cr 0&-2i\end{pmatrix}\oplus \begin{pmatrix}{4i\over 3}&0\cr 0&-{2i\over 3}\end{pmatrix}\otimes 1_3,\cr
&&\tau_L=-i1_2\oplus{i\over 3}1_2\otimes 1_3,\cr
\mbox{for }t_{W}^a:&&\tau_R=0,\cr
&&\tau_L=i\sigma^a\oplus i\sigma^a\otimes 1_3 , a=1,2,3\cr
\mbox{ for }t_{C}^a:&&\tau_R=\tau_L=0\oplus 1_2\otimes i {\lambda^a},  a=1,\ldots,8 
\end{eqnarray*}
and where we choose the bases $\sigma^a$ and $\lambda^a$ of Pauli and Gell-Mann matrices, normalized by $\Tr(\sigma^a\sigma_b)=\Tr(\lambda^a\lambda_b)=2\delta^a_b$. Formula  \eqref{Fluctuationdevelopee} is just a decomposition of $D_\omega$ on a particular basis chosen to recognize the usual fields. But one notices an intruder, namely the $X$-field. It has to be set to zero by hand: this is the infamous unimodularity problem (see Ref.~\onlinecite{Suijlekom}, chap 8 for a thorough exposition) which affects all NCG models of particle physics to date. The removal of the $X$ field, which is equivalent to anomaly freeness,  is consistent with \eqref{gt} only if we restrict $u$ to have determinant $1$, yielding the correct gauge group $U(1)\times SU(2)\times SU(3)$.

The computation\cite{Besnard-19-2} of \eqref{genCL} yields (for $N=3$ generations):
\begin{eqnarray}
n\calL_b&=&-40g^2B_{\mu\nu}B^{\mu\nu}-24g_w^2W_{\mu\nu}^aW^{\mu\nu}_a-24g_s^2G_{\mu\nu}^aG^{\mu\nu}_a\cr
&&+16A|D_\mu H|^2-8V_0(|H|^2-1)^2
\end{eqnarray}
where $H$ is the second column of the quaternion $q$, and 
\begin{eqnarray}
D_\mu H&=&(\partial_\mu+\frac{1}{2}ig_w W_\mu^a\sigma_a+\frac{1}{2}igB_\mu)H,
\end{eqnarray}
from which  we see that the doublet $H$ has hypercharge $1$.

The constants $A$ and $V_0$  can be computed from the entries of $D_F$. More precisely, under the \emph{genericity hypothesis} that $\Upsilon$ is invertible and that any matrix commuting with both $\Upsilon_\nu \Upsilon_\nu^\dagger$ and $\Upsilon_e\Upsilon_e^\dagger$ (resp. $\Upsilon_u \Upsilon_u^\dagger$ and $\Upsilon_d \Upsilon_d^\dagger$) is scalar, we find that 
\begin{multline}
A=\Tr(\Upsilon_e\Upsilon_e^\dagger+\Upsilon_\nu \Upsilon_\nu^\dagger+3\Upsilon_u\Upsilon_u^\dagger+3\Upsilon_d\Upsilon_d^\dagger)\cr
V_0= \|\widetilde{ \Upsilon_\nu \Upsilon_\nu^\dagger}\|^2+\|\widetilde{ \Upsilon_e \Upsilon_e^\dagger}\|^2+3\|\widetilde{\Upsilon_u\Upsilon_u^\dagger}\|^2+3\|\widetilde{\Upsilon_d \Upsilon_d^\dagger}\|^2 \cr 
+2{\|\widetilde{\Upsilon_e\Upsilon_e^\dagger}\|^2\|\widetilde {\Upsilon_\nu \Upsilon_\nu^\dagger}\|^2\over \|\widetilde{\Upsilon_\nu \Upsilon_\nu^\dagger}-\widetilde{\Upsilon_e \Upsilon_e^\dagger}\|^2}\sin^2\theta_\ell+6{\|\widetilde{\Upsilon_u\Upsilon_u^\dagger}\|^2\|\widetilde {\Upsilon_d \Upsilon_d^\dagger}\|^2\over \|\widetilde{\Upsilon_u \Upsilon_u^\dagger}-\widetilde{\Upsilon_d \Upsilon_d^\dagger}\|^2}\sin^2\theta_q,  \label{aV0}
\end{multline}
where the angles $\theta_\ell$ and $\theta_q$ are defined up to sign by
\begin{eqnarray}
\Re\Tr(\widetilde{\Upsilon_\nu \Upsilon_\nu^\dagger}\widetilde{\Upsilon_e \Upsilon_e^\dagger})=\|\widetilde{\Upsilon_\nu \Upsilon_\nu^\dagger}\|\|\widetilde{\Upsilon_e \Upsilon_e^\dagger}\|\cos(\theta_\ell)\cr
\Re\Tr(\widetilde{\Upsilon_u \Upsilon_u^\dagger}\widetilde{\Upsilon_d \Upsilon_d^\dagger})=\|\widetilde{\Upsilon_u \Upsilon_u^\dagger}\|\|\widetilde{\Upsilon_d \Upsilon_d^\dagger}\|\cos(\theta_q),
\end{eqnarray}
the norm of a matrix $A\in M_N(\bbC)$ is  the Hilbert-Schmidt norm $\|A\|=\sqrt{\Tr(A^\dagger A)}$ and $\tilde A=A-\frac{\Tr(A)}{N}1_N$.

{\small {\bf Remark:} The   tildes and the sine terms are not present in the traditional formalism of Euclidean Connes-Lott theory. Their presence can be traced back to the use of the $J$-symmetrized background. 
}

In order to normalize gauge kinetic terms as usual one has to set the coupling constants to the special values
\begin{equation}
g_w^2=g_s^2=\frac{5}{3}g^2=\frac{n}{96}.\label{gcunif}
\end{equation}
Similarly we introduce the Higgs field
\begin{equation}
\phi=4\sqrt{\frac{A}{n}}H,\label{defHiggs}
\end{equation}
so that the kinetic term is $|D_\mu \phi|^2$. This fixes the values of the Higgs quartic coupling to
\begin{equation}
\lambda=\frac{V_0n}{32a^2}.\label{lambdaunif}
\end{equation} 
Since the Higgs potential is obtained via \eqref{genCL} as a square, its minimum   is zero and the Higgs field does not contribute to the cosmological constant. Moreover the minimum is obtained for $|H|=1$, the vev of $\phi$ therefore satisfies
\begin{equation}
\frac{v}{\sqrt{2}}=4\sqrt{\frac{A}{n}}.\label{vunif}
\end{equation}
Let us now look at the fermionic action. It is given by
\begin{equation}
S_f(D_\omega,\Psi)=\frac{1}{2}(\Psi,D_\omega \Psi)\label{fermionicLag}
\end{equation}
where $\Psi$ belongs to the subspace $\calK_{\rm Phys}$ of $\calK$ determined by \eqref{MWconditions}. Alternatively\cite{Besnard-19-4}, one can take for $\Psi$ a generic element of $\calK$ and use the action 
\begin{equation}
S_f(D_\omega,\Psi)'=\frac{1}{2}(\pi\Psi,D_\omega \pi\Psi)
\end{equation}
where $\pi=\frac{(1+J)(1+\chi)}{4}$ is the projector on $\calK_{\rm Phys}$. Either way we obtain all the usual terms of the SM. Let us compute some of them in order to show the peculiarities of the calculations of a NCG model. This will also yield the precise interpretation of the matrices $\Upsilon$ and $M$ entering $D_F$.  Seeing $\Psi$ as a field with values in $S\otimes \calK_F$, one sees that
\begin{equation}
\Psi=\sum_p\psi_R^p\otimes p_R+J_M\psi_R^p\otimes p_R^c+\psi_L^p\otimes p_L-J_M\psi_L^p\otimes p_L^c\label{solutionMW}
\end{equation} 
where $p$ runs through the orthonormal basis of elementary fermions. One then just has to plug \eqref{Fluctuationdevelopee} and \eqref{solutionMW} into \eqref{fermionicLag}. The result for the gauge term in the electron sector is for instance
\begin{equation}
2(e_R,\gamma^\mu e_R)(\frac{1}{2}g B_\mu)+(e_L,\gamma^\mu e_L)(\frac{1}{2}g B_\mu)
\end{equation}
which can be used to check the consistency of the charge assignments and convention for the covariant derivative. Now the Yukawa and Majorana terms are 
\begin{multline}
\frac{1}{2}(\Psi,1\hat\otimes(D_F+\Phi(q-1)+\Phi(q-1)^o) \Psi)= \cr
\sum_{i,i'}\big((\nu_L^{i},\alpha\nu_R^{{i'}})(\Upsilon_\nu)_{ii'}+(\nu_L^{ i},\beta e_R^{{i'}})(\Upsilon_e)_{ii'}-(e_L^{i},\beta^*\nu_R^{{i'}})(\Upsilon_\nu)_{ii'}\cr
 +(e_L^{i},\alpha^* e_R^{{i'}})(\Upsilon_e)_{ii'} 
 +(u_L^{i},\alpha u_R^{{i'}})(\Upsilon_u)_{ii'}+(u_L^{i},\beta d_R^{{i'}})(\Upsilon_d)_{ii'}\cr
 -(d_L^{i},\beta^*u_R^{{i'}})(\Upsilon_u)_{ii'}+(d_L^{i},\alpha^*d_R^{{i'}})(\Upsilon_d)_{ii'}\big)\cr
 +\frac{1}{2}\sum_{i,i'}( J_M\nu_R^{i},\nu_R^{{i}'}) (m)_{ii'}
 +h.c.
\end{multline}
where we recall  $H=\begin{pmatrix}
\beta\cr
\alpha^*
\end{pmatrix}$ is related to the Higgs field $\phi=\begin{pmatrix}
\phi^+\cr
\phi^0
\end{pmatrix}$ by \eqref{defHiggs}. However since the minimum of the Higgs potential corresponds to $q=1$ by construction, we see that $\Upsilon_\nu,\Upsilon_e,\Upsilon_u,\Upsilon_d$ are exactly the Dirac mass matrices of fermions. We also see that  $m/2$ is the Majorana mass matrix of right-handed neutrinos.

{\small {\bf Remark:} From \eqref{aV0} one can then infer the following interpretation for $A$ and $V_0$: $A$ is the sum of the Dirac masses of fermions squared, while the first four terms of $V_0$ are variances of fermion masses. 
}

The unification of the gauge couplings   which is predicted by the approach holds at some energy scale $\mu_{\rm unif}$.  At this energy the bosonic Lagrangian is given by \eqref{genCL}, and the prediction \eqref{gcunif}, \eqref{lambdaunif} and \eqref{vunif} are supposed to hold.  There is also a relation between the $W$-bosons and fermions masses at the unification scale.   Charge eigenstates $W_\mu^\pm$ are introduced as usual and their tree-level mass is 
\begin{equation}
m_W={1\over 2}vg_w.\label{massW}
\end{equation}
From this we obtain:
\begin{eqnarray}
m_W^2&=&{1\over 4}v^2g_w^2\cr
&=&{1\over 4}{n\over 96}32n \Tr(\Upsilon_e \Upsilon_e^\dagger+\Upsilon_\nu \Upsilon_\nu^\dagger+3\Upsilon_u^\dagger \Upsilon_u+3\Upsilon_d^\dagger \Upsilon_d)\cr
&=&{1\over 12}\sum\mbox{ squared masses of fermions},
\end{eqnarray}
where in the second line we have used \eqref{vunif}. In particular   we obtain the bound
\begin{equation}
m_t\le 2m_W.
\end{equation}
We note that this prediction is different from the one obtained with the spectral action, which is\cite{Dungen-12} $m_{t}\le \sqrt{8/3}m_W$.

We will suppose as in Ref.~\onlinecite{Connes-Chamseddine-06} that only one Dirac neutrino mass $m_D$ is non-negligible with respect to the  top quark mass. Introducing the couplings $y_t$ and $y_\nu$ defined such that the Dirac masses of the top quark and   neutrino are 
\begin{eqnarray}
m_t=\frac{1}{\sqrt{2}}y_tv,& &m_D=\frac{1}{\sqrt{2}}y_\nu v,\label{diracmasses}
\end{eqnarray}
one obtains from \eqref{vunif} and \eqref{aV0}
\begin{eqnarray*}
v^2&\approx&\frac{32}{n}(3m_{t}^2+m_D^2),
\end{eqnarray*}
where we have neglected all masses except for $m_t$ and $m_D$. Plugging in \eqref{diracmasses} and introducing 
\begin{equation}
\rho:=y_\nu/y_t,
\end{equation}
we get the relations
\begin{equation}
y_t=\sqrt{\frac{n}{16(3+\rho^2)}}, y_\nu=\rho\sqrt{\frac{n}{16(3+\rho^2)}} \label{ytunif}
\end{equation}
In order to obtain   $\lambda$ at unification scale, let us observe that for a matrix $A\in M_N(\bbC)$, one has $\Tr(\tilde A^2)=\Tr(A^2)-{1\over N}\Tr(A)^2$. If the spectrum of  $A$ is dominated by an eigenvalue $M^2$ we  have the approximation $\Tr(A^2)\simeq \Tr(A)^2\simeq M^4$, so that
\begin{equation}
\Tr(\tilde A^2)\simeq {N-1\over N}M^4 
\end{equation}
Using this observation and equations \eqref{aV0} and \eqref{lambdaunif},   we obtain the approximation
\begin{equation}
\lambda=\frac{n(N-1)(3+\rho^4)}{32N(3+\rho^2)^2}.\label{lambdaunif2}
\end{equation}
Together with \eqref{gcunif}, \eqref{ytunif} and \eqref{lambdaunif2} can be used as initial values\footnote{Compare with formulas on p. 95 of Ref.~\onlinecite{Dungen-12}.} for a run down of the renormalization group equations to the experimentally accessible energy scales. We see that there are two free  parameters $n$ and $\rho$, and a  starting energy $\mu_{\rm unif}$, which is tied to $n$ by \eqref{gcunif}. Since the gauge couplings never exactly come together under the SM RGE, we cannot define a precise value for $\mu_{\rm unif}$. Instead we will use the following strategy: we give a value to $n$ and set $\mu_{\rm unif}$ so as to minimize the error on gauge couplings at the $Z$ mass scale. To estimate this error we use the relative standard deviation
\begin{equation}
RSD_g=\frac{\sqrt{\scriptstyle{3((g(m_Z)-g)^2+(g_w(m_Z)-g_w)^2+(g_s(m_Z)-g_s)^2)}}}{\scriptstyle{g+g_w+g_s}},
\end{equation}
where $g_i(m_Z)$ are the running coupling obtained by running down the RGE from energy scale $\mu_{\rm unif}$, and $g_i$ are the experimental values at $m_Z$. We keep only the values of $n$ for which $RSD_g$ can be lowered to less that 5 percent. This leaves the interval $[22,35]$ for $n$, corresponding to energies going from $10^{12}$ GeV to the Planck scale. The relative standard deviation of gauge couplings will then be (at 1-loop) the same  function of $n$   for all the models studied in this paper.

Once $\mu_{\rm unif}$ is found for a given $n$, we set $\rho$ so as to minimize the RSD of the top and Higgs masses computed according to the formula:
\begin{equation}
RSD_{m}=\frac{\sqrt{\scriptstyle{2((m_t(172)-172)^2+(m_h(125)-125)^2})}}{\scriptstyle{172+125}}.
\end{equation}
Note that in order to measure how well the model fits the experimental data we do not use the predicted  pole masses   of the top and Higgs but the running masses computed at their respective experimental values (for instance, the top mass in table \ref{SMresults} is the running mass $m_t(172)$). The figures should be close if the model is good.  Note also that we calculated $m_t$ from \eqref{diracmasses} and the Higgs mass from
\begin{equation}
m_h^2=2\lambda v^2,\label{higgsmassSM}
\end{equation}
using  $v=246.66$ GeV  and ignoring renormalization of $v$. We explored the parameter space using $1$-loop RGE and found no value of $\rho$ for which  $RSD_{m}$  were less than ten percent. Worse, the Higgs mass is always off by more than 30 percent. Example of results are shown in table \ref{SMresults}. Note that if instead of calculating the RSD of masses we fix $\rho$ to   fit the top mass perfectly (which happens for $\rho\approx 1.5$), then the Higgs mass is way too large (close to 170  GeV).

One could argue that instead of \eqref{gcunif} one should use as initial values for the gauge couplings those  which are run up from the Z scale, setting $RSD_g$ to $0$ by construction (this is the strategy used in Refs.~\onlinecite{Chamseddine-06} and  \onlinecite{Dungen-12}). Hence, at one loop, the values of $g_w,g_s$ and $g$  at $\mu_{\rm unif}$ are switched to
\begin{eqnarray*}
\bar g&=&(g(m_Z)^2-2kb_1(\log(\mu_{\rm unif})-\log(m_Z)))^{-1/2}\cr
\bar g_w&=&(g_w(m_Z)^2-2kb_2(\log(\mu_{\rm unif})-\log(m_Z)))^{-1/2}\cr
\bar g_s&=&(g_s(m_Z)^2-2kb_3(\log(\mu_{\rm unif})-\log(m_Z)))^{-1/2},
\end{eqnarray*}
where $k=1/16\pi^2$, $b_1=41/6$, $b_2=-19/6$, $b_3=-7$, and the couplings at $m_Z$ are the experimental values.  This can be justified by embedding the SM  in a larger (unspecified) extension  with a threshold happening just at $\mu_{\rm unif}$. Hence some threshold correction $\delta$ would change $g$ into $g+\delta=\bar g$ and so forth. However this does not change the results significantly as far as $RSD_m$ is concerned (see table \ref{SMresults2}). We conclude that the model is not satisfactory for empirical  reasons. We will see in  section \ref{BLextSM} that it also suffers from a theoretical inconsistency.


\begin{table}[hbtp]
\begin{tabular}{|c|c|c|c|c|c|c|c|}
\hline 
$n$     & $24$   & $26$ & 28 & $30$ & $32$ \\ 
\hline 
$\log_{10}(\mu_{\rm unif}/{\rm GeV})$      & ${18.27}$   & $16.75$ & 15.45 & $14.32$ & $13.34$ \\ 
\hline 
$\rho$   & $1.84$   & $1.85$ & $1.85$ &  $1.86$ & $1.86$ \\ 
\hline
$RSD_g$    & $0.032$   & $0.020$ & 0.015 & $0.020$ & $0.031$ \\  
\hline 
$RSD_m$    & $0.131$   & $0.141$ & 0.151 &  $0.162$ & $0.173$\\ 
\hline 
Top mass   & 161  & $162$ & 163   & 163  &  164 \\
\hline
Higgs mass  & $161$   & $164$ & 168   &  172 & 175  \\
\hline
\end{tabular}
\caption{Relative standard deviations of gauge couplings at $m_Z$ energy scale. RSD of top and Higgs masses computed at scales $172$ and $125$ GeV respectively. $\mu_{\rm unif}$ chosen to minimize  $RSD_g$  and $\rho$  to minimize  $RSD_m$.}\label{SMresults}
\end{table}

\begin{table}[hbtp]
\begin{tabular}{|c|c|c|c|c|c|c|c|}
\hline 
$n$     & $24$   & $26$ & 28 & $30$ & $32$ \\ 
\hline 
$\log_{10}(\mu_{\rm unif}/{\rm GeV})$      & ${18.27}$   & $16.75$ & 15.45 & $14.32$ & $13.34$ \\ 
\hline 
$\rho$   & $1.09$   & $1.07$ & 1.07 &  $1.03$ & $1.02$ \\ 
\hline
$RSD_m$    & $0.076$   & $0.077$ & 0.077 &  $0.078$ & $0.079$\\ 
\hline 
Top mass   & $159$  & $160$ & 161 & $161$ & $161$ \\
\hline
Higgs mass  & $139$   & $141$ & 141 & $143$ & $143$ \\
\hline
\end{tabular}
\caption{Same as table \ref{SMresults} when the gauge couplings are corrected by an unknown threshold effect.}\label{SMresults2}
\end{table}

\section{The  Chamseddine-Connes model}\label{CCmodel}

In Ref.~\onlinecite{Chamseddine-12} Connes and Chamseddine performed the RGE analysis at one loop of the Euclidean NCSM with the Spectral Action, extended with a real scalar added by hand. The general agreement with the experimental values of the top and Higgs masses was an important step for the subsequent development of the theory. For this reason, we are going to briefly reanalyze this model and use it as a benchmark. We do not enter into any detail, refering instead to Refs.~\onlinecite{Chamseddine-12}  or   \onlinecite{Dungen-12}.

Connes and Chamseddine set themselves in the case where there is effectively only one familly of right-handed neutrinos. Their model depends on two parameters: the unification scale $\mu_{\rm unif}$ and $\rho:=y_t/y_\nu$. They found that for any $\mu_{\rm unif}$ in the allowed zone there exists a $\rho$ which gives a good fit for the Higgs mass. However, they remarked that for such a $\rho$ the predicted top mass was off by a few percents and argued  that the 2-loop effects could correct this. Instead of fitting the Higgs mass first and looking at the error on the top mass, we give in table \ref{CCresults2} the interval to which $\rho$ must belong in order to obtain  $RSD_m<0.05$, as well as the minimum of $RSD_m$ when $\rho$ varies in this interval. We see that the agreement with experimental values is never much better that 5 $\%$.   However, the mass of the real scalar has been estimated\cite{Dungen-12}  to be at least of the order $10^{12}$ GeV. Hence it must decouple from the RGE under this energy, giving rise to a threshold effect.  We see from tables \ref{CCresults3} and \ref{CCresults5} that the model gives better results ($RSD_m$ around $3\%$) when this threshold effect is taken into account.

\begin{table}[hbtp]
\begin{tabular}{|c|c|c|c|c|c|c|c|}
\hline 
$\log_{10}(\mu_{\rm unif}/{\rm GeV})$      & ${18.27}$   & $16.75$ & 15.45     \\ 
\hline
$RSD_m<0.05$   & $[1.38,1.47]$    & $[1.34;1.39]$     & $[1.31,1.32]$     \\ 
\hline
best fit & 0.048  & 0.049  & 0.050   \\
\hline
\end{tabular}
\caption{Results of the run down of the Chamseddine-Connes model with no threshold.}\label{CCresults2}
\end{table}

\begin{table}[hbtp]
\begin{tabular}{|c|c|c|c|c|c|c|c|}
\hline 
$\log_{10}(\mu_{\rm unif}/{\rm GeV})$      & ${18.27}$   & $16.75$ & 15.45     \\ 
\hline
$RSD_m<0.05$   & $[1.42,1.62]$    & $[1.38,1.54]$     & $[1.34,1.48]$     \\ 
\hline
best fit & 0.031  & 0.032  & 0.033   \\
\hline
\end{tabular}
\caption{Results of the run down of the Chamseddine-Connes model with a threshold at $10^{12}$ GeV under which the real scalar decouples.}\label{CCresults3}
\end{table}

\begin{table}[hbtp]
\begin{tabular}{|c|c|c|c|c|c|c|c|}
\hline 
$\log_{10}(\mu_{\rm unif}/{\rm GeV})$      & ${18.27}$   & $16.75$ & 15.45     \\ 
\hline
$RSD_m<0.05$   & $[1.42,1.59]$    &  $[1.37,1.52]$   &   $[1.35,1.47]$      \\ 
\hline
best fit &  0.024 & 0.026  & 0.028   \\
\hline
\end{tabular}
\caption{Results of the run down of the Chamseddine-Connes model with a threshold at $10^{14}$ GeV for the real scalar.}\label{CCresults5}
\end{table}
%

\section{The B-L-extended NC Standard Model }\label{BLextSM}
\subsection{Formulation of the model and prediction of the couplings at high energy}
The model presented in section \ref{SMback} is not consistent with the viewpoint of algebraic backgrounds since with the latter it can be shown that the B-L symmetry must be gauged\cite{Besnard-19-3}. In order to fulfil this requirement we  extend the algebra by a factor of $\bbC$. The \emph{extended} SM triple and background $\calS_{\rm SM}^{\rm ext}$ and $\calB_{\rm SM}^{\rm ext}$, respectively, are hence defined exactly as before except for the following modifications:
\begin{itemize}
\item $\calA_F$ is replaced with $\calA_F^{\rm ext}=\bbC\oplus\bbC\oplus\bbH\oplus M_3(\bbC)$,
\item $\pi_F$ is replaced with $\pi_F^{\rm ext}$ defined by
\begin{equation}
\pi_F^{\rm ext}(\lambda,\lambda',q,a)={\rm diag}(\tilde q_\lambda,\tilde q,1_2\otimes(\lambda'\oplus a)\otimes 1_N, 1_2\otimes(\lambda'\oplus a)\otimes 1_N)\label{SMextrep}
\end{equation}
\item $\Omega^1_F$ is replaced with $(\Omega^1_F)^{\rm ext}$ containing the $1$-forms
\begin{equation}
\omega_F^{\rm ext}=\begin{pmatrix}
0&\Upsilon^\dagger \tilde q_1&z_1M^\dagger&0\cr \tilde q_2 \Upsilon&0&0&0\cr z_2M&0&0&0\cr 0&0&0&0
\end{pmatrix}, q_1,q_2\in \bbH, z_1,z_2\in\bbC.\label{SMext1forms}
\end{equation}
\end{itemize} 
Let us note that $\calB_{\rm SM}^{\rm ext}$ only satisfies weak $C_1$. The choice of \eqref{SMext1forms} for the   bimodule of finite $1$-forms is such that $D_F$ is still a regular Dirac operator for the extended background. This can serve as a first justification, but it shoud be noted that $(\Omega^1_F)^{\rm ext}$ can also be found as a solution of the following two constraints\cite{bes-20-b}: the extended background must 1) satisfy weak $C_1$, and 2) satisfy $C_1$ when the algebra elements are restricted to $\calA_F$. Moreover this solution is almost unique: $\Upsilon$ and $M$ must have the form given by equations \eqref{Y} and \eqref{formeM}, the only remaining freedom beeing in the form \eqref{formeYell}.

Within the extended background we can fluctuate around the same Dirac operator as before, or make the simpler choice $D=D_M\hat\otimes 1$ (for the same result). With the latter choice, the elements of the extended configuration space are\footnote{The presence or absence of $\frac{1}{2}$ factors in front of the abelian fields is dictated by our choice of generators $t_Y$ and $t_{B-L}$. Our conventions match Ref.~\onlinecite{Langacker}.}
\begin{eqnarray}
D+i\gamma^\mu\hat\otimes (Xt_X+{1\over 2}gB_\mu t_Y+{1\over 2}g_wW^a_\mu t_W^a+{1\over 2}g_sG^a_\mu t_C^a&&\cr
+g_{Z'}Z_\mu't_{B-L}+1\hat\otimes(\Phi(q)+\Phi(q)^o+\sigma(z))&&\label{extFluctuationdevelopee}
\end{eqnarray} 
with $z$ a complex field,
\begin{equation}
\sigma(z)=\begin{pmatrix}
0&0&z^*M^\dagger&0\cr 0&0&0&0\cr zM&0&0&0\cr 0&0&0&0
\end{pmatrix},
\end{equation}
and the B-L generator is $t_{B-L}={\rm diag}(\tau_R,\tau_L,\tau_R^*,\tau_L^*)\otimes 1_N$ with
\begin{equation}
\tau_R=\tau_L=-i1_2\oplus\frac{i}{3}1_2\otimes 1_3.
\end{equation}
Inserting \eqref{extFluctuationdevelopee} into \eqref{genCL} yields (for $N=3$)
\begin{eqnarray}
n{\cal L}_b &=&-40g^2B_{\mu\nu}B^{\mu\nu}-24g_w^2W_{\mu\nu a}W^{\mu\nu a}-24g_s^2G_{\mu\nu a}G^{\mu\nu a}\cr
&&-64g_{Z'}^2Z_{\mu\nu}'{Z'}^{\mu\nu}-{64}gg_{Z'}Z_{\mu\nu}'B^{\mu\nu}\cr
&&+16A|D_\mu H|^2+8B|D_\mu z|^2-8V_0(|H|^2-1)^2\cr
&&-8W_0(|z|^2-1)^2-16K(|H|^2-1)(|z|^2-1)\label{extboslag}
\end{eqnarray}
where $A$ and $V_0$ keep the same meaning as in \eqref{aV0} and the other constants are
\begin{eqnarray}
b&=&\Tr(m^\dagger m)\cr
W_0&=&\|\widetilde {mm^\dagger}\|^2,\cr
K&=&\Re\Tr(\widetilde{\Upsilon_\nu^\dagger \Upsilon_\nu}\widetilde{m^\dagger m}),
\label{parameters}
\end{eqnarray}
The covariant derivative of $z$ is
\begin{equation}
D_\mu z=(\partial_\mu+2ig_{Z'}Z_\mu')z
\end{equation}
which shows that $z$ has B-L charge $2$. Normalizing the gauge kinetic terms to
\begin{equation}
-{1\over 4}|B_{\mu\nu}|^2 -{1\over 4}|W_{\mu\nu}^a|^2-{1\over 4}|G_{\mu\nu}^a|^2-{1\over 4}|Z_{\mu\nu}'|^2-{\kappa \over 2}Z_{\mu\nu}'{B}^{\mu\nu},
\end{equation}
we obtain
\begin{eqnarray}
g_w^2=g_s^2={5\over 3}g^2={8\over 3}g_{Z'}^2={n\over 96},& \kappa=\sqrt{2\over 5}\label{gaugekin}
\end{eqnarray}
To deal with the kinetic mixing term we perform the standard triangular transformation\footnote{Mathematically, this is the unique upper Cholevski transformation which diagonalizes the quadratic form \eqref{gaugekin}.}  to obtain new fields $\tilde B$ and $\tilde Z'$ 
\begin{equation}
\begin{pmatrix} B\cr Z'\end{pmatrix}= \begin{pmatrix}1&-{\kappa\over \sqrt{1-\kappa^2}}\cr 0&{1\over \sqrt{1-\kappa^2}}\end{pmatrix}\begin{pmatrix}\tilde B\cr \tilde Z'\end{pmatrix}.\label{triang}
\end{equation}
Rewriting the covariant derivative (or Dirac operator) in terms of the tilded fields introduce new couplings constants $g'$ and $\tilde{g}$ defined such that
\begin{equation}
\frac{1}{2}gB_\mu t_Y+g_{Z'}Z_\mu't_{B-L}=\frac{1}{2}g\tilde B_\mu t_Y+\tilde Z_\mu'(\frac{1}{2}\tilde gt_Y+g't_{B-L}),
\end{equation}
which yields
\begin{eqnarray}
g'=\frac{g_{Z'}}{\sqrt{1-\kappa^2}},&\tilde{g}=-\frac{\kappa g}{\sqrt{1-\kappa^2}}\label{gprimgtilde}
\end{eqnarray}
From \eqref{gaugekin} we obtain $\frac{\tilde{g}}{g'}=-\frac{4}{5}$ which exactly the same value as that coming from $SO(10)$ unification\cite{accomando}.

Let us turn to the scalar sector. Introducing the normalized Higgses
\begin{eqnarray}
\phi=4\sqrt{\frac{A}{n}}H,&\xi=\sqrt{\frac{8B}{n}}z,
\end{eqnarray}
the scalar Lagrangian becomes $|D_\mu \phi|^2+|D_\mu \xi|^2-V(\phi,\xi)$, where the potential is
\begin{equation}
V(\phi,\xi)=m_1^2|\phi|^2+m_2^2|\xi|^2+\lambda_1 |\phi|^4+\lambda_2|\xi|^4+\lambda_3|\phi|^2|\xi|^2+\mu 
\end{equation}
with
\begin{gather}
\lambda_1=\frac{V_0 n}{32 A^2},\lambda_2={W_0n\over 8B^2}, \lambda_3={Kn\over 8 AB}\cr
m_1^2=-{V_0+K\over A}, m_2^2=-2{W_0+K\over B},\cr
\mu=8\frac{V_0+W_0+2K}{n} \label{couplings2}
\end{gather}
Once again, the minimum of the potential is obtained for $|z|=|H|=1$ directly from \eqref{extboslag}, and this gives the vev's:
\begin{equation}
v=4\sqrt{\frac{2A}{n}},\ v'=4\sqrt{\frac{B}{n}}.\label{vevs}
\end{equation}
The fermionic action gets new terms coming from the $Z'$ and $z$-fields. The latter is 
\begin{equation}
\frac{1}{2}z\sum_{i,i'}(\bar\nu^i_R,\nu^{i'}_R)m_{ii'}+h.c.
\end{equation}
and we recover the Majorana mass matrix $\frac{1}{2}m$ when $z=1$.

Let us now look for a set of initial conditions for the  RGE. In order to do that we have to choose a scenario for the hierarchy of Majorana masses. In this respect it is interesting to note  from \eqref{couplings2} and \eqref{parameters} that the initial value of $\lambda_2$ and $\lambda_3$ are proportional to $\|\widetilde {mm^\dagger}\|^2$ and $\|\widetilde {mm^\dagger}\|$ respectively. This means that $m$ cannot be too close to the identity matrix since in that case $\lambda_2$ and $\lambda_3$ would be only radiatively generated and be too small to have a sizable impact on the Higgs mass (this argument is supported by numerical simulations). We will thus use the opposite scenario in which one Majorana mass $m_M$  dominates the others. As far as the Dirac masses are concerned, we will continue to suppose as in section \ref{NCSM} that one Dirac mass $m_D$ is dominant and non-negligible with respect to the top mass.

Let us define the Majorana and Dirac Yukawa coupling matrix $Y_N$ and $Y_\nu$ by
\begin{eqnarray}
m&=&\sqrt{2}Y_Nv',\cr
\Upsilon_\nu&=&\frac{1}{\sqrt{2}} Y_\nu v.\label{mtauM}
\end{eqnarray} 
Performing a unitary change of basis in the left and right neutrino spaces, we can always suppose that $\Upsilon_\nu$ is diagonal, and thus of the form $\Upsilon_\nu={\rm diag}(m_D,0,0)$. In the same basis we have $Y_\nu={\rm diag}(y_\nu,0,0)$ and $y_\nu$ is related to $n$ and $\rho$ by the same formula \eqref{ytunif} as in section \ref{NCSM}.

Now from \eqref{parameters} we obtain
\begin{equation}
B\approx (m_M)^2\label{approxb}
\end{equation}
and from \eqref{vevs} we find
\begin{equation}
m=4m_M\sqrt{\frac{2}{n}}Y_N\label{relym}
\end{equation}
Since $m$ is symmetric, its singular value decomposition can be written
\begin{equation}
m=U\Sigma U^T\label{msdv}
\end{equation}
where $U$ is a unitary matrix. The fact that $\Sigma\approx {\rm diag}(m_M,0,0)$ will allow us to suppress many degrees of freedom in $U$, retaining only its first column $(a,b,c)^T$.   Thus we have 
\begin{equation}
m\approx m_M \begin{pmatrix}
a^2&ab&ac\cr
ba&b^2&bc\cr
ca&cb&c^2
\end{pmatrix}
\end{equation}
and the only remaining freedom we have is to   multiply $U$ to the right by an orthogonal matrix commuting with $\Sigma$, yielding a global change of sign of $(a,b,c)$. Now from \eqref{relym} we obtain
\begin{equation}
Y_N\approx \frac{1}{4}\sqrt{\frac{n}{2}}\begin{pmatrix}
a^2&ab&ac\cr
ba&b^2&bc\cr
ca&cb&c^2
\end{pmatrix},\label{ytauminit}
\end{equation}
where $(a,b,c)$ is a point of the complex $2$-sphere, uniquely defined up to a global sign.  Using the spherical coordinates with origin $(1,0,0)$  we can parametrize it by
\begin{eqnarray}
a&=&\cos\theta  e^{i\alpha},\cr
b&=&\sin\theta\cos\varphi e^{i\beta},\cr
c&=&\sin\theta\sin\varphi e^{i\gamma}
\end{eqnarray}
with $\alpha\in [0,\pi[$, $\beta,\gamma\in]-\pi,\pi]$, $\theta,\varphi\in[0,\pi/2[$.

To obtain the other initial values we observe that we can write
\begin{eqnarray}
W_0&\approx& \frac{2}{3}(m_M)^4,\cr
K&\approx& \frac{3|a|^2-1}{3}\rho^2 m_t^2(m_M)^2,
\end{eqnarray}
where $K$ is obtained by direct computation. From \eqref{couplings2} we then  obtain the initial conditions
\begin{align}
\lambda_1=\frac{(3+\rho^4)n}{48(3+\rho^2)^2}, \lambda_2=\frac{n}{12}, \lambda_3= \frac{(3|a|^2-1)\rho^2n}{24(3+\rho^2)}.\label{lambdaInit}
\end{align}
 Equations \eqref{ytunif}, \eqref{gaugekin},  \eqref{ytauminit} and \eqref{lambdaInit} now form a complete set of initial values\footnote{Note that the perturbativity bounds are satisfied for all the admissible values of $n$ and for all $\rho$ and $a$.} for the RGE. The latter, which we have computed with the Pyr@te 3 software\cite{pyrate}  are given in appendix \ref{appendixRGE}. When the run down is performed we obtain predictions for the top quark mass and the lightest scalar of the model, which is identified to the SM Higgs. The formula for the top mass is the same as in the SM, but the mass of the   Higgs gets a correction. Indeed,  the masses of the two Higgses  satisfy\cite{Basso-10,Coriano-16}:
\begin{equation}
m_{h_1/h_2}^2=\lambda_1v^2+\lambda_2{v'}^2\mp\sqrt{(\lambda_1v^2-\lambda_2{v'}^2)^2+(\lambda_3vv')^2}\label{higssmasses}
\end{equation}
Using $v^2\ll v'$ (see equation \eqref{vprimebound} below) one finds
\begin{equation}
m_{h_1}^2\approx 2v^2\left(\lambda_1-\frac{\lambda_3^2}{4\lambda_2}\right),\label{HiggsmassBL}
\end{equation}
which replaces equation \eqref{higgsmassSM}. In the next sections we present the results of the run down and compare the predictions of the top and SM Higgs mass with their experimental values.

\subsection{The gauge couplings}
\begin{figure}[hbtp]\label{gaugecouplingsrundown}
\includegraphics[scale=0.45]{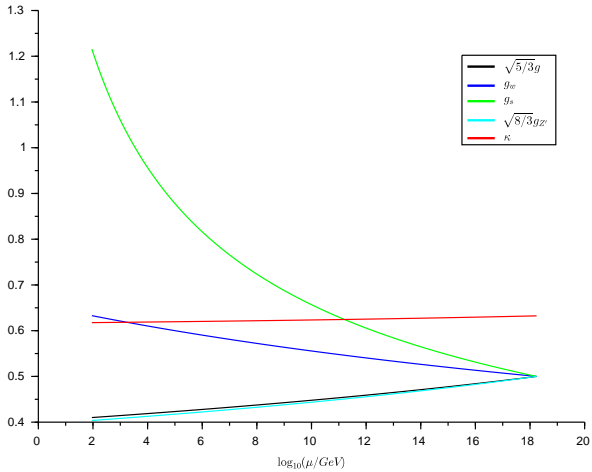}
\caption{The normalized gauge couplings $\sqrt{5/3}g,g_w,g_s,\sqrt{8/3}g_{Z'}$ and $\kappa$ as functions of $\log_{10}(\mu/{\rm GeV})$ for $n=24$.}
\end{figure}

The running of the gauge couplings  presents interesting peculiarities (figure \ref{gaugecouplingsrundown}) which need to be discussed.

One notes the extreme stability of $\kappa$ and the almost perfect equality of the two normalized abelian couplings at all scales. An explanation of the first phenomenon is the following: let us change the $(t_Y,t_{B-L},t_W^a,t_C^b)$ basis to an orthogonal one. This is done just by removing the orthogonal projection of $t_{B-L}$ onto $t_Y$, defining the new basis vector
\begin{eqnarray}
t_{Z'}&:=&t_{B-L}-(t_{B-L},t_Y){t_Y\over \|t_Y\|^2}\cr
&=&t_{B-L}-{2\over 5}t_Y\label{changebases}
\end{eqnarray} 
Since the curvature is linear in the abelian fields, this change of basis also removes the kinetic mixing, and is equivalent to \eqref{triang} at the level of fields components. Though \eqref{triang} is meaningful in any B-L model and scale-dependent, \eqref{changebases} is scale-independent. This shows that the value $\kappa=\sqrt{2\over 5}$ is stable under RGE. This feature is shared by any theory in which the gauge couplings are unified, as shown in Ref.~\onlinecite{Aguila-88}. It is also proved in the latter paper that the normalized couplings of the abelian fields associated with the diagonalizing basis, here $\sqrt{5/3}g_1$ and $\sqrt{8/3}g_{Z'}$, are equal at all scales since their beta functions coincide. By \eqref{gprimgtilde} and the constancy of $\kappa$, we obtain that $g'$ and $\tilde g$ are also equal at all scales when correctly normalized.

\subsection{The Yukawa and scalar sectors in first approximation}\label{case24}
In this section we present the results of the running down of the RGE with the initial conditions \eqref{ytunif}, \eqref{ytauminit}, and \eqref{lambdaInit} with $a=1,b=c=0$. This amounts to consider only one species of right neutrinos (which by definition will be the $\tau$). Note also that    we will not consider any threshold correction. We hence make the same simplifying assumptions as in the Chamseddine-Connes model. Although crude, this approximation will allow us to get a feel of the general properties of the B-L extension.  In particular,   it can be seen from table \ref{BLresults} that, in stark constrast with the NC SM,  compatibility with the experimental values of the top and   Higgs masses can  be achieved for any allowed value of $n$, as long as $\rho$ satisfies
\begin{equation}
1.35\le \rho\le 1.58
\end{equation}
These bounds are roughly the same as in the Chamseddine-Connes model, but the $RSD_m$ is much improved, as can be seen by comparing tables \ref{CCresults2} and \ref{BLresults}.

We can also obtain some information on the masses at experimentally accessible scales. At any energy $\mu$ we have the relations
\begin{eqnarray}
m_\tau^M&=&\sqrt{2}y_\tau^M(\mu)v'(\mu)\cr
m_\tau^D(\mu)&=&\frac{1}{\sqrt{2}}y_\nu(\mu)v(\mu)
\end{eqnarray}
From the see-saw  formula $m_{\rm light}\approx (m_\tau^D)^2/m_\tau^M$ we also get
\begin{equation}
m_{\rm light}(\mu)\approx \frac{y_\nu(\mu)^2v(\mu)^2}{2\sqrt{2}y_\tau^M(\mu)v'(\mu)}
\end{equation}
For all the  values of $n$ and $\rho$   which are allowed by the experimental values of the SM gauge couplings and top and Higgs masses, we have (see table \ref{BLresults})
\begin{equation}
y_\tau^M(m_Z)\approx y_\nu(m_Z)\approx 0.5
\end{equation}
and with $v=246.66$ GeV (neglecting the running of $v$ from the Fermi scale) this yields
\begin{equation}
m_{\rm light}(m_Z)\approx \frac{  10^4}{v'(m_Z)}
\end{equation}
Now from the bound $m_{\rm light}\le 0.2$ eV at the $Z$ scale on light neutrino masses\cite{PDG-19}  and  one obtains 
\begin{equation}
v'(m_Z)\gtrsim 5\times 10^{14}\mbox{ GeV}\label{vprimebound}
\end{equation}
This makes $v'/v$ very large. Consequently  the mixing angle $\theta'$ which rotates to  the mass eigenstates of the $Z$ and $Z'$-bosons, given by \cite{Coriano-16}
\begin{equation}
\tan(2\theta')={2\tilde g\sqrt{g_w^2+g_Y^2}\over \tilde g^2+16{g'}^2\left({v'\over v}\right)^2-g_w^2-g_Y^2}
\end{equation}
is vanishingly small at the $Z$-scale. In this regime the $Z'$-boson mass is given by (Ref.~\onlinecite{Coriano-16}, formula (46))
\begin{equation}
M_{Z'}(m_Z)\simeq 2{g'}(m_Z){v'}(m_Z)\gtrsim 3.7\times 10^{14}\mbox{ GeV} 
\end{equation}
and from \eqref{HiggsmassBL}
\begin{equation}
m_{h_2}(m_Z)\simeq \sqrt{2\lambda_2(m_Z)}v'(m_Z) \gtrsim 4\times 10^{14}\mbox{ GeV}
\end{equation}
These values are of course well out of reach of accelerators, so that we obtain without surprise compatibility with the LEP bounds\cite{DELPHI-95,cacciapaglia}
\begin{eqnarray}
|\theta'|&<&10^{-3}\cr
\frac{M_{Z'}}{g'}&>&7\mbox{ TeV}
\end{eqnarray}
The masses of the $Z'$ and the new scalar show that the model should get a threshold correction at $10^{14}$ GeV. This will be done in section \ref{fullcpx} below.


\begin{table}[hbtp]
\begin{tabular}{|c|c|c|c|c|c|}
\hline 
$n$     & $24$   & $26$ & 28 & $30$   \\ 
\hline 
$\log_{10}(\mu_{\rm unif}/{\rm GeV})$      & ${18.27}$   & $16.75$ & 15.45 & $14.32$   \\ 
\hline 
best $RSD_m$ & 0.001  & 0.008   & 0.017 & 0.024      \\
\hline
$\rho_{best}$  & 1.47  & 1.46   & 1.44 & 1.43      \\
\hline
$RSD_m<0.05$   & $[1.35,1.58]$   & $[1.35,1.56]$ & $[1.35,1.53]$ &  $[1.35,1.51]$  \\ 
\hline
$y_\nu(m_Z)$   &  $0.50$    & $0.53$  & $0.56$ & $0.59$   \\
\hline
$y_\tau^M(m_Z)$   &  $0.48$    & $0.51$  & $0.53$  &  $0.55$   \\
\hline
\end{tabular}
\caption{The sixth line shows the intervals of $\rho$ for which $RSD_m$ falls below 5 percent. The two last lines show the values of $y_\nu$ and $y_\tau^M$ at the $Z$-scale for $\rho=1.5$.}\label{BLresults}
\end{table}
 
One notes that the allowed interval for $\rho$ tends to narrow down as $n$ grows, and that the agreement becomes less and less good. The best fit is obtained for $n=24$, which is a particularly interesting  value\footnote{There is of course no reason other than aesthetical to limit ourselves to integer values of $n$.}  since it is   the dimension of $\calK_0$ and could be interpreted   as a natural normalization of the trace in \eqref{genCL}. We find it remarkable that this value which is the most aesthetically appealing not only falls in the allowed range but yields the best fit for the Higgs and top masses ! This also sets all gauge couplings to $1/2$ and corresponds to the   energy $10^{18.27}$ GeV, which is quite close to the Planck scale.  

\begin{figure}[hbtp]
\includegraphics[scale=0.45]{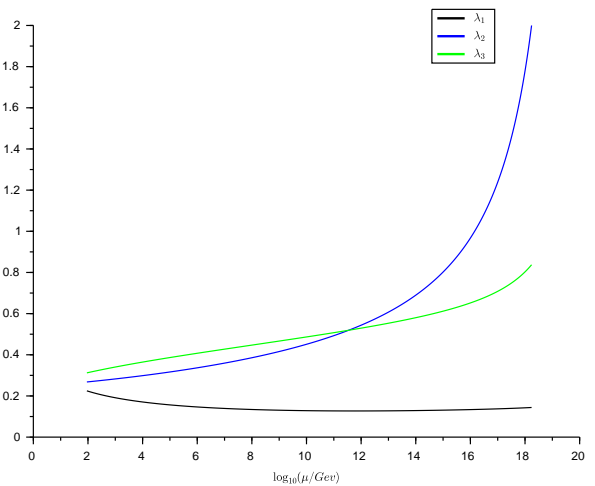}
\caption{Running of the quartic couplings for $n=24$, $\rho=1.47$.}\label{perturbativity}
\end{figure}

\begin{figure}
\includegraphics[scale=0.45]{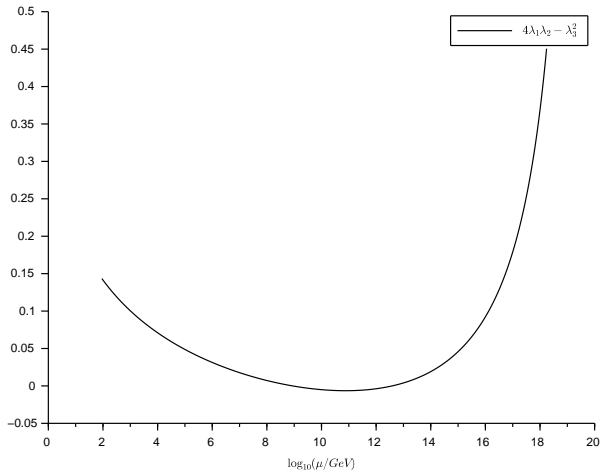}
\caption{The running of $\Delta=4\lambda_1\lambda_2-\lambda_3^2$ for $n=24$ and $\rho=1.47$. Though this is difficult to spot, $\Delta$ does fall below $0$ around $10^{11}$ GeV.}\label{positivity}
\end{figure}

Besides compatibility with experimental values, there are also two interesting theoretical constraints which are 
  the perturbativity bounds\cite{accomando}
\begin{equation}
\lambda_{1,2,3}<\sqrt{4\pi}
\end{equation}
and the stability bounds
\begin{equation}
\lambda_{1,2}>0,\ \Delta:=4\lambda_1\lambda_2-\lambda_3^2>0.
\end{equation}
The first one is satisfied at all scales and for all values of $(n,\rho)$ already allowed by the experimental constraints (see figure \ref{perturbativity} for an example of running of the quartic couplings). The stability bounds are satisfied at all scales and for all $n$ as soon as $\rho\le 1.45$. It is intriguing that the value   $\rho=1.45$ for which the minimum of $\Delta$ is   exactly $0$ is so close to the value fitting best the Higgs and top masses which is $1.47$. Note however that $\rho$ has to be $>1.5$ in order for $\Delta$ to take on negative values of non-negligible magnitude (see \ref{positivity} for a plot with $\rho=1.47$).

\begin{figure}[hbtp]
\includegraphics[scale=0.45]{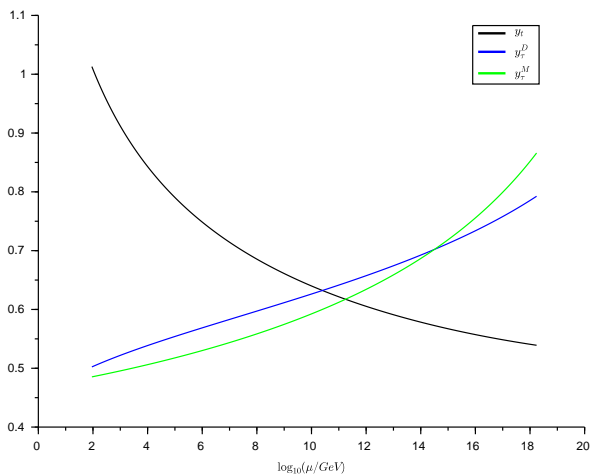}
\caption{Running of the Yukawa couplings for $n=24$, $\rho=1.47$.}
\end{figure}

\subsection{The full parameter space and threshold corrections}\label{fullcpx}
In this section we explore the full parameter space for $Y_N$ and we also include the threshold corrections.

Up to now we ran down one and the same RGE, defined in the MS scheme, from the GUT scale to the Z scale. The latter procedure is  suspicious because of the   Appelquist-Carazzone decoupling theorem, which is not manifest in the  mass independent MS renormalization scheme. Hence  the couplings to the very massive $\xi$, $Z'$ and $\nu_R$  particles are to be suppressed by hand when the energy scale falls below a threshold which, according to the previous analysis, should be at least of the order $10^{14}$ GeV.  
 
The 1-loop RGE are thus supplemented by the tree-level   matching conditions\cite{Elias-12}:
\begin{eqnarray}
\lambda&=&\lambda_1-\frac{\lambda_3^2}{4\lambda_2},\label{matching}
\end{eqnarray}
which can be obtained from the continuity of the Higgs mass (formulas \eqref{higgsmassSM} and \eqref{HiggsmassBL}). Since there is no shift in the $W$ mass, and in the limit $v^2\ll v'$ the shift in the $Z^0$ mass is negligible, the matching conditions for $g_1$ and $g_2$ are trivial.

Up to a set of Lebesgue measure zero, the parameter space $\calY$ for $Y_N$ is the product of the spherical positive octant $\calO$ (containing the moduli $(|a|,|b|,|c|)$ and parametrized by $(\theta,\varphi)$) with a 3-torus $\calT$ of phases $(\alpha,\beta,\gamma)$. Let us now explore this parameter space.

We start with the numerical observation that  $(\alpha,\beta,\gamma)$ have almost no impact on the Higgs, top quark and  light neutrinos masses.  To see this we first fix a value of $\rho$ and a point in $\calO$ and run down the RGE for a random sample of $10^3$ points in $\calT$ equipped with the uniform law. We then compute the standard deviations of $RSD_m$ and  the light neutrino mass $m_\ell$ for this sample. The results are displayed in table \ref{phasedep}.

\begin{table}[hbtp]
\begin{tabular}{|c|c|c|c|c|c|c|c|}
\hline  
$(\rho,\theta,\varphi)$   & $(2,0.1,0.2)$       & $(2.1,0.7,0.8)$       \\ 
\hline
st. dev. of $RSD_m$& $3.8\times 10^{-5}$  & $6.8\times 10^{-5}$          \\
\hline
st. dev. of $m_\ell$ (in eV) &  $2.2\times 10^{-5}$ & $1.3\times 10^{-4}$  \\
\hline 
\end{tabular}
\caption{This table exemplifies the near independence of $RSD_m$ and $m_\ell$ on the complex phases $(\alpha,\beta,\gamma)$ for $10^3$ random points in the 3-torus and some fixed values of $(\rho,\theta,\varphi)$. The energy scale is set to $10^{18.27}$ GeV, corresponding to $n=24$.}\label{phasedep}
\end{table}

We can understand this behaviour from the 1-loop RGE (see appendix \ref{appendixRGE}). First we see that $Y_t$ and $Y_\nu$ do not depend on $Y_N$ at all. As for the quartic couplings, we see that they depend on $Y_N$ only through $\tr(Y_NY_N^*)$, $\tr((Y_NY_N^*)^2)$, and $\tr(Y_\nu Y_N^*Y_NY_\nu^\dagger)$. However these 3 traces are independent on the phases at $\mu_{\rm unif}$, so that the dependency only appears indirectly from radiative corrections which will stay small between $\mu_{\rm unif}$ and the threshold at $10^{14}$ GeV where they are set to zero. From now on we will consider only the case $\alpha=\beta=\gamma=0$ for simplicity.


\begin{table}[hbtp]
\begin{tabular}{|c|c|c|c|c|c|c|c|}
\hline  
$(n,\rho,\theta)$   & $(24,2,0.1)$       & $(26,2,0.5)$   & $(28,2.3,1)$    \\ 
\hline
best $RSD_m$& 0.03081   & 0.10336      & 0.10632     \\
\hline
worst $RSD_m$&  0.03083  &  0.10338     & 0.10633    \\
\hline
st. dev. of $m_\ell$ (in eV) &  $3.7\times 10^{-6}$ & $1.1\times 10^{-5}$ &   $7.4\times 10^{-6}$ \\
\hline 
\end{tabular}
\caption{This table shows the near independence of $RSD_m$ and $m_\ell$ on the longitude $\varphi$ for $100$ random points  with $(n,\rho,\theta)$ fixed.}\label{phidep}
\end{table}

As it shows on table \ref{phidep}, $RSD_m$ is also nearly independent of $\varphi$.  This again can be understood from the RGE. Let $\delta$ be an angle and $R_\delta$ be the rotation matrix $R_\delta=\begin{pmatrix}
1&0&0\cr 0&\cos \delta&-\sin\delta\cr 0&\sin\delta&\cos\delta
\end{pmatrix}$.  Clearly $R_\delta$ commutes with $Y_\nu$ at $\mu_{\rm unif}$, and one has 
\begin{eqnarray}
R_\delta U\Sigma U^T R_\delta^{-1}&=&R_\delta U\Sigma U^T R_\delta^T\cr
&=&(R_\delta U)\Sigma (R_\delta U)^T
\end{eqnarray}
from which we infer that $R_\delta Y_N(a,b,c) R_\delta^{-1}=Y_N(a',b',c')$, where $(a',b',c')^T$ is the image of $(a,b,c)^T$ by the rotation $R_\delta$. Let us replace $Y_\nu$ with $R_\delta Y_\nu R_\delta^{-1}=Y_\nu$ and $Y_N$ with $R_\delta Y_N(a,b,c) R_\delta^{-1}=Y_N(a',b',c')$ at the unification scale. Since $a=a'$ this does not change the initial conditions for the quartic couplings. Let us call $Y_\nu'(\mu)$ and $Y_N'(\mu)$ the   running matrices with these new conditions. Since the beta functions for $Y_N$ and $Y_\nu$ are covariant with respect to  unitary changes of basis,   at any energy scale we have $Y_\nu'(\mu)=R_\delta Y_\nu(\mu) R_\delta^{-1}$ and $Y_N'(\mu)=R_\delta Y_N(\mu)R_\delta^{-1}$. Now the beta functions for the quartic couplings are invariant under the change $(Y_\nu,Y_N)\rightarrow (Y_\nu',Y_N)$, and since the initial conditions are the same, we obtain that the scalar masses are invariant under   rotations of the vector $(a,b,c)^T$ around the first axis.  The conclusion we can draw from this study is that among the 5-dimensional parameter space of $Y_N$ only $|a|=\cos\theta$ is really relevant.  This parameter has a direct interpretation in terms of the  matrices entering $D_F$ through the formula
\begin{equation}
|a|^2=\frac{|\bra\Upsilon_\nu  ,m\ket|}{\|\Upsilon_\nu\|\|m\|}:=\cos\epsilon.\label{defepsilon}
\end{equation}
which defines the angle $\epsilon$ between the matrices $\Upsilon_\nu$ and $m$. Because of this interpretation we will now express the results in terms of $\epsilon$ rather that $\theta$. Table \ref{epsilonrho} shows  the minimum, maximum and best fit values of $\epsilon$  as functions of $\rho$ for $n=24$. We see that $\epsilon_{\rm min}$ which is constantly equal to zero at first starts growing rapidly when $\rho\approx 2.15$ and meets $\epsilon_{\rm max}$ between $\rho=2.25$ and $\rho=2.3$ so that no value of $\epsilon$ is accepted for larger values of $\rho$. This behaviour is similar for the other values of $n$, with a lower value of $\epsilon_{\rm max}$, so that  the model predicts the bound
\begin{equation}
\epsilon<0.22
\end{equation}
The intervals of $\rho$ for which there exists an accepted value of $\epsilon$ are given in table \ref{randomsearch2} for different values of $n$. They yield the prediction of the model for the parameter $\rho$, namely
\begin{equation}
1.88\le \rho\le 2.36.
\end{equation}
We also see from this table that as $n$ grows the starting energy is closer to the threshold, and the effects of the new fields become less important, resulting in a worse fit to the experimental values. At the limit when $n=30$, we are almost at the threshold and we see that no value of $\rho$ is accepted. 

\begin{table}[hbtp]
\begin{tabular}{|c|c|c|c|c|c|c|c|c|}
\hline 
$\rho$     &  1.95  & 2 & 2.05 & 2.1 & 2.15 & 2.2& 2.25 & 2.3  \\ 
\hline 
$\epsilon_{\rm min}$      & 0   & 0 & 0  & 0 & 0.04 & 0.12 & 0.19& $\emptyset$ \\ 
\hline
$\epsilon_{\rm best}$   &0     &  0 &  0.03  &  0.10 & 0.14 & 0.17&0.19 & $\emptyset$  \\ 
\hline
$\epsilon_{\rm max}$ &  0.13 & 0.17   & 0.19  & 0.21 & 0.22 & 0.22& 0.20 & $\emptyset$   \\
\hline
best $RSD_m$ & 0.034 & 0.027 & 0.028 & 0.034 & 0.039 & 0.044 & 0.050 & 0.055  \\
\hline
\end{tabular}
\caption{$\epsilon_{\rm min,max}$ are respectively the minimum and maximum values of $\epsilon$ for which $RSD_m<0.05$. $\epsilon_{\rm best}$ is the value of $\epsilon$ for which $RSD_m$ is minimized. Here $n=24$.}\label{epsilonrho}
\end{table}



\begin{table}[hbtp]
\begin{tabular}{|c|c|c|c|c|c|c|c|}
\hline 
$n$     & $24$   & $26$ & 28 & $30$  \\ 
\hline 
$\log_{10}(\mu_{\rm unif}/{\rm GeV})$      & ${18.27}$   & $16.75$ & 15.45  & 14.32 \\ 
\hline
$RSD_m<0.05$   & $[1.88,2.25]$      &  $[1.99,2.35]$ &  $[2.12,2.36]$  &  $\emptyset$  \\ 
\hline
best fit & 0.0267 & 0.0290   & 0.0354  &   \\
\hline
$\rho_{best}$ & 2.02 & 2.13 & 2.23 &  \\
\hline
Higgs mass (in GeV) & 127.8 & 126.4 & 129.7 &  \\
\hline
Top mass (in GeV) & 166.7 & 165.9 & 165.3 &  \\
\hline
\end{tabular}
\caption{Interval of $\rho$ for which there exists $\epsilon$ such that $RSD_m<0.05$. The masses of the Higgs and top are given for the best fit parameters. Threshold at $10^{14}$ GeV.}\label{randomsearch2}
\end{table}

In table \ref{randomsearch3} we also display the results obtained when we correct the initial values of the SM gauge coupling with a threshold effect, as described in section \ref{NCSM}. We see that  this   improves $RSD_m$ for $n=24$ but does not affect much the allowed values for $\rho$.

\begin{table}[hbtp]
\begin{tabular}{|c|c|c|c|c|c|c|c|}
\hline 
$n$     & $24$   & $26$ & 28 & $30$  \\ 
\hline 
$\log_{10}(\mu_{\rm unif}/{\rm GeV})$      & ${18.27}$   & $16.75$ & 15.45 & $14.32$   \\ 
\hline
$RSD_m<0.05$   & $[1.91,2.35]$      &  $[1.98,2.35]$ &  $[2.10,2.37]$  &  $\emptyset$  \\ 
\hline
best fit & 0.019  &   0.027  & 0.037   &   \\
\hline
$\rho_{best}$ &  2.05 & 2.13  &  2.23 & \\  
\hline
\end{tabular}
\caption{Same as table \ref{randomsearch2} except that the gauge couplings are corrected by a threshold effect.}\label{randomsearch3}
\end{table}

{\small {\bf Remark:} We see that the model is quite sensitive to neutrino physics, and it would be interesting to connect it with some parameters on which there exist experimental bounds, like the entries of the PMNS matrix. 

In the flavour basis the Dirac and Majorana mass matrices are $\Upsilon_\nu^f$ and $m^f$. By unitary change of bases in the Left and Right neutrino spaces  we can diagonalize $\Upsilon_\nu^f$. Since we assumed that $\Upsilon_\nu$ were diagonal, we can thus write
\begin{eqnarray}
\Upsilon_\nu&=&V_L^\dagger \Upsilon_\nu^f V_R,\cr
m&=&V_R^Tm_fV_R.
\end{eqnarray}
From \eqref{msdv} we thus obtain $m_f=V_R^*U\Sigma U^TV_R^\dagger$. Using the see-saw approximation, the masses of the light neutrinos are then the singular values of $\calM=\Upsilon_\nu^f (m^f)^{-1}(\Upsilon_\nu^f)^T$. Here we have
\begin{eqnarray}
\calM&=&V_L\Upsilon_\nu V_R^\dagger (V_RU^*\Sigma^{-1}U^\dagger V_R^T)V_R^*\Upsilon_\nu V_L^T\cr
&=&V_L\Upsilon_\nu U^*\Sigma^{-1}U^\dagger\Upsilon_\nu V_L^T\cr
&:=&V_L \Delta V_L^T
\end{eqnarray}
where $\Delta$ is diagonal and we identify $V_L$ as the PMNS matrix. We see that $\cos\epsilon$ can be written as
\begin{equation}
\cos\epsilon=\frac{|\bra V_L^\dagger \Upsilon_\nu^f  , V_R^Tm^f\ket|}{\|\Upsilon_\nu^f\|\|m^f\|} 
\end{equation}
}

\begin{figure}[hbtp]
\includegraphics[scale=0.4]{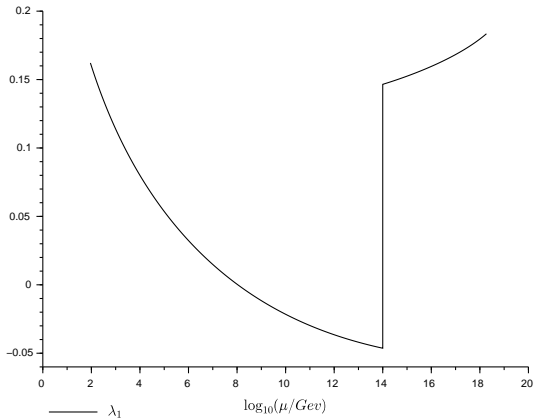}
\caption{Running of $\lambda_1$ for $n=24$, $\rho=1.9$, $\epsilon=0$.}\label{instabthresh}
\end{figure}

Let us conclude this section with the observation that the field $\xi$, decoupling above the electoweak vacuum instability scale $\sim 10^8$ GeV, is not able to cure this problem anymore. As can be seen on figure \ref{instabthresh}, $\lambda_1$  takes small negative values just below the threshold. The figure is drawn for particular values of the parameters, but the  phenomenon is general.

\subsection{2-loop effects}
When we use the beta functions at 2 loops we must first lower a little the unification energy as a function of $n$ to have the best fit of the gauge couplings at the $Z$ energy. The results of the run down in the first approximation regime (no threshold, $a=1,b=c=0$) is then quite similar to the one found in section \ref{case24} as can be seen from table \ref{BLresults2Loops}. The fit of the top and Higgs masses is a little less good but stays of the same order as the fit of the gauge couplings, which is enough. In order to take into account the 2-loop effects in the more general regime, we should have to go to 1-loop matching conditions, which goes beyond the scope of this paper. However, we expect from the present section that these corrections would be small in front of the effects induced by the threshold itself and by the parameter $\epsilon$.

\begin{table}[hbtp]
\begin{tabular}{|c|c|c|c|c|c|}
\hline 
$n$     & $24$   & $26$ & 28  \\ 
\hline 
$\log_{10}(\mu_{\rm unif}/{\rm GeV})$     & ${18.07}$   & $16.57$ & 15.28\\  
\hline 
$RSD_g$  & 0.033 &  0.020   &  0.013         \\
\hline
best $RSD_m$ & 0.011  & 0.018   & 0.026      \\
\hline
$\rho_{best}$  & 1.53  & 1.52   & 1.51      \\
\hline
$RSD_m<0.05$   & $[1.41,1.64]$   & $[1.42,1.62]$ & $[1.42,1.59]$  \\ 
\hline
$m_t$   &  $174.2$    & $175.8$  & $177.3$     \\
\hline
$m_{\rm Higgs}$   &  $124.4$    & $142.0$  & $122.7$      \\
\hline
\end{tabular}
\caption{Results of the run down of the 2-loop RGE.}\label{BLresults2Loops}
\end{table}

\section{Conclusion}

The 1-loop RGE analysis of NCG particle models had already been performed in the Euclidean context followed by a Wick rotation using the spectral action \cite{Chamseddine-07-2} or Connes-Lott action \cite{Martin-98} (without right-handed neutrinos), but never before in a genuinely Lorentzian framework as we did in this paper. Despite different initial conditions, our conclusions were similar to the ones of  these previous studies, with in particular a large discrepancy between the predicted and experimental values of the Higgs boson mass.  In order to remedy this situation, and also to comply with the new framework of algebraic backgrounds, we introduced a B-L-extension of the NC Standard Model, based on the algebra $\bbC\oplus\bbC\oplus \bbH\oplus M_3(\bbC)$ and a generalization of the Connes-Lott action. We explored the RG flow of this model, including the  corrections introduced by 1) the decoupling of the particles with very high masses, and  2) the relative positions of the Majorana and Dirac neutrino mass matrices. We found these corrections to be important (and dominant with respect to 2-loop effects) and that there exists a region of the parameter space compatible with the experimental values of the top quark, Higgs boson, and light neutrino masses. Hence, the  model we analyzed is the first one coming from NCG which   1) is Lorentzian right from the start, 2) is consistent with the  algebraic background point of view, 3) yields  masses for the top quark and Higgs boson which agree well with the experimental values (improving significantly on the Chamseddine-Connes model in this respect). The model is predictive, since  one has to start from a rather small region of the parameter space to fit the top and Higgs masses. The predicted parameters are the quotient of the Yukawa couplings $y_\nu$ and $y_t$, as well as the angle between the Dirac and Majorana mass matrices for neutrinos. Another prediction of the model is that the Majorana masses of the heavy neutrinos are not too close together, thus ruling out the scenario of a universal Majorana coupling which is sometimes considered in the literature \cite{Coriano-16}. It would be interesting to compare these predictions with the ones coming from the spectral action in Euclidean signature. This will be the subject of a forthcoming paper.

Let us conclude by saying that the model we have explored in this paper is incomplete, as can be seen for instance from the running of the gauge couplings. A possible theoretical development would be to include the finite algebra in the Clifford algebra of a $10$-dimensional space, with possible connections with $SO(10)$ GUT. From the phenomenological point of view, the most pressing issue would be to relate the parameters in the neutrino sector which are predicted by the model we studied with those accessible to current experiments.

\section{Acknowledgments}

The authors would like to thank  Koen van den Dungen, Benjamin Fuks, Julia Harz, Thomas Krajewski, Carlo Marzo, Walter van Suijlekom for kindly answering our questions, Mark Goodsell for very useful comments and suggestions, and Lohan Sartore for his invaluable help with the Pyr@te 3 software. 

\section{Data availability}
The Pyr@te 3 model files and Scilab codes used in this paper are available upon request to the authors.

\appendix
\section{Renormalization group equations the B-L extended model}\label{appendixRGE}
We use the standard notation
\begin{equation*}
\beta\left(X\right) \equiv \mu \frac{d X}{d \mu}\equiv\frac{1}{\left(4 \pi\right)^{2}}\beta^{(1)}(X).
\end{equation*}
Down quarks and electrons Yukawa couplings are neglected. The Yukawa coupling matrix of up quarks and neutrinos are  {$Y_d$, $Y_\nu$} and $Y_N$, the latter being associated with the Majorana mass term. {The RGE are obtained from  Pyr@te 3\cite{pyrate}}.

\subsection{Gauge couplings}
{\allowdisplaybreaks

\begin{align*}
\begin{autobreak}
\beta^{(1)}(g) =\frac{41}{6} g^{3}
\end{autobreak}
\end{align*}
\begin{align*}
\begin{autobreak}
\beta^{(1)}({g'}) =

+ 12 {g'}^{3}

+ \frac{41}{6} {g'} \tilde{g}^{2}

+ \frac{32}{3} {g'}^{2} \tilde{g}
\end{autobreak}
\end{align*}
\begin{align*}
\begin{autobreak}
\beta^{(1)}(g_{2}) =- \frac{19}{6} g_{2}^{3}
\end{autobreak}
\end{align*}
\begin{align*}
\begin{autobreak}
\beta^{(1)}(g_{3}) =-7 g_{3}^{3}
\end{autobreak}
\end{align*}
\begin{align*}
\begin{autobreak}
\beta^{(1)}(\tilde{g}) =

+ \frac{32}{3} g^{2} {g'}

+ \frac{32}{3} {g'} \tilde{g}^{2}

+ \frac{41}{3} g^{2} \tilde{g}

+ 12 {g'}^{2} \tilde{g}

+ \frac{41}{6} \tilde{g}^{3}
\end{autobreak}
\end{align*}
}

\subsection{Yukawa couplings}
{\allowdisplaybreaks

\begin{align*}
\begin{autobreak}
\beta^{(1)}(Y_u) =

+ \frac{3}{2} Y_u Y_u^{\dagger} Y_u

+ 3 \tr\left(Y_u Y_u^{\dagger} \right) Y_u

+ \tr\left(Y_\nu Y_\nu^{\dagger} \right) Y_u

-  \frac{17}{12} g^{2} Y_u

-  \frac{2}{3} {g'}^{2} Y_u

-  \frac{5}{3} {g'} \tilde{g} Y_u

-  \frac{17}{12} \tilde{g}^{2} Y_u

-  \frac{9}{4} g_{2}^{2} Y_u

- 8 g_{3}^{2} Y_u
\end{autobreak}
\end{align*}

\begin{align*}
\begin{autobreak}
\beta^{(1)}(Y_\nu) =

+ \frac{3}{2} Y_\nu Y_\nu^{\dagger} Y_\nu

+ 2 Y_\nu Y_N^{*} Y_N

+ 3 \tr\left(Y_u Y_u^{\dagger} \right) Y_\nu

+ \tr\left(Y_\nu Y_\nu^{\dagger} \right) Y_\nu

-  \frac{3}{4} g^{2} Y_\nu

- 6 {g'}^{2} Y_\nu

- 3 {g'} \tilde{g} Y_\nu

-  \frac{3}{4} \tilde{g}^{2} Y_\nu

-  \frac{9}{4} g_{2}^{2} Y_\nu
\end{autobreak}
\end{align*}
\begin{align*}
\begin{autobreak}
\beta^{(1)}(Y_N) =

+ Y_\nu^{\trans} Y_\nu^{*} Y_N

+ Y_N Y_\nu^{\dagger} Y_\nu

+ 4 Y_N Y_N^{*} Y_N

+ 2 \tr\left(Y_N Y_N^{*} \right) Y_N

- 6 {g'}^{2} Y_N
\end{autobreak}
\end{align*}
}

\subsection{Quartic couplings}
{\allowdisplaybreaks

\begin{align*}
\begin{autobreak}
\beta^{(1)}(\lambda_1) =

+ 24 \lambda_1^{2}

+ \lambda_3^{2}

- 3 g^{2} \lambda_1

- 3 \tilde{g}^{2} \lambda_1

- 9 g_{2}^{2} \lambda_1

+ \frac{3}{8} g^{4}

+ \frac{3}{4} g^{2} \tilde{g}^{2}

+ \frac{3}{4} g^{2} g_{2}^{2}

+ \frac{3}{8} \tilde{g}^{4}

+ \frac{3}{4} g_{2}^{2} \tilde{g}^{2}

+ \frac{9}{8} g_{2}^{4}

+ 12 \lambda_1 \tr\left(Y_u Y_u^{\dagger} \right)

+ 4 \lambda_1 \tr\left(Y_\nu Y_\nu^{\dagger} \right)

- 6 \tr\left(Y_u Y_u^{\dagger} Y_u Y_u^{\dagger} \right)

- 2 \tr\left(Y_\nu Y_\nu^{\dagger} Y_\nu Y_\nu^{\dagger} \right)
\end{autobreak}
\end{align*}
\begin{align*}
\begin{autobreak}
\beta^{(1)}(\lambda_2) =

+ 20 \lambda_2^{2}

+ 2 \lambda_3^{2}

- 48 {g'}^{2} \lambda_2

+ 96 {g'}^{4}

+ 8 \lambda_2 \tr\left(Y_N Y_N^{*} \right)

- 16 \tr\left(Y_N Y_N^{*} Y_N Y_N^{*} \right)
\end{autobreak}
\end{align*}
\begin{align*}
\begin{autobreak}
\beta^{(1)}(\lambda_3) =

+ 12 \lambda_1 \lambda_3

+ 8 \lambda_2 \lambda_3

+ 4 \lambda_3^{2}

-  \frac{3}{2} g^{2} \lambda_3

- 24 {g'}^{2} \lambda_3

-  \frac{3}{2} \tilde{g}^{2} \lambda_3

-  \frac{9}{2} g_{2}^{2} \lambda_3

+ 12 {g'}^{2} \tilde{g}^{2}

+ 6 \lambda_3 \tr\left(Y_u Y_u^{\dagger} \right)

+ 2 \lambda_3 \tr\left(Y_\nu Y_\nu^{\dagger} \right)

+ 4 \lambda_3 \tr\left(Y_N Y_N^{*} \right)

- 16 \tr\left(Y_\nu Y_N^{*} Y_N Y_\nu^{\dagger} \right)
\end{autobreak}
\end{align*}
}

\section{Notation}

 \begin{table}[h]
\begin{tabular}{|c|c|}
\hline
$\calA$  & allowed part of $\calY$ \\ 
\hline
$(a,b,c)^T$  & first column of the unitary matrix diagonalizing $m$ \\ 
\hline
$\alpha,\beta,\gamma$  & arguments of $a,b,c$ \\ 
\hline
$\mu_{\rm unif}$  & unification energy scale \\ 
\hline 
$n$  & normalization constant of the  Connes-Lott  action \\ 
\hline
$\calO$ & space of $(\theta,\varphi)$ (spherical positive octant) \\ 
\hline 
$\rho$  & $y_\nu/y_t$ (at unification scale) \\ 
\hline
$\calT$ & space of $(\alpha,\beta,\gamma)$ ($3$-torus) \\ 
\hline 
$(\theta,\varphi)$ & latitude and longitude of $(|a|,|b|,|c|)$ \\ 
\hline
$\calY$ &   space of $(a,b,c)$ ($=\calO\times\calT$) \\ 
\hline
\end{tabular}
\caption{Free parameters} 
\end{table}

  \begin{table}
\centering
\begin{tabular}{|c|c|c|c|c|}
\hline
    & $q_s$ & $q_w$ & $q_Y$ & $q_{Z'}$ \\
\hline
$\nu_{R}$ & 0 & 0 & 0 & -1 \\
$e_R$ & 0 & 0 & -2 & -1 \\
$u_R$ & 1 & 0 & 4/3 & 1/3 \\
$d_R$ & 1 & 0 & -2/3 &  1/3 \\
$\ell_L$ & 0 & 1 & -1 & -1 \\
$q_L$ & 1 & 1 & 1/3 & 1/3 \\
$\phi$ & 0 & 1 & 1 & 0 \\
$\xi$ & 0 & 0 & 0 & 2 \\ \hline
\end{tabular}
\caption{Charges of fermions and scalars in the B-L extended model, where $q_s$ is the SU(3) strong charge,
$q_w$ the SU(2) weak charge, $q_Y$ the U(1) hypercharge and $q_{Z'}$ the U(1) B-L charge. The covariant derivative of each
field is $
D_\mu = \partial_\mu + \frac12 q_s g_s G^a_\mu \lambda_a +
 \frac12 q_w g_w W^a_\mu\sigma_a
+ \frac12 q_Y i g_Y Y_\mu + q_{Z'} i g_{Z'} Z'_\mu$,
where $\lambda_a$ are the Gell-Mann matrices and $\sigma_a$ the Pauli matrices.
\label{Tablecharge}}
\end{table}

\bibliographystyle{apsrev}
\bibliography{qed}
\end{document}